\documentclass[12pt,a4paper,english,pdftex]{revtex4}

\usepackage{amssymb}
\usepackage{amsmath,graphicx,color,epsfig}
\usepackage{pstricks}
\usepackage{float}

\newcommand\nn{\nonumber}
\newcommand\ba{\begin{eqnarray}}
\newcommand\ea{\end{eqnarray}}

\newcommand{\br}[1]{\left( #1 \right)}
\newcommand{\brs}[1]{\left[ #1 \right]}
\newcommand{\brf}[1]{\left\{ #1 \right\}}

\begin{document}

\begin{flushright}
  DESY 11-003\\
FTUAM-11-038\\
March 2011
\end{flushright}

\title{Improved Sensitivity to Charged Higgs Searches in Top Quark Decays
$t \to b H^+ \to b (\tau^+ \nu_\tau)$  at the LHC using $\tau$ Polarisation and
Multivariate Techniques}

\author{Ahmed~Ali}
\email{ahmed.ali@desy.de}
\affiliation{Deutsches Elektronen-Synchrotron DESY, D-22607 Hamburg, Germany}

\author{Fernando Barreiro}\email{fernando.barreiro@uam.es}
\author{Javier Llorente}\email{Javier.Llorente.Merino@cern.ch}
\affiliation{Universidad Autonoma de Madrid (UAM),
Facultad de Ciencias C-XI, Departamento de Fisica,
Cantoblanco, Madrid 28049, SPAIN }


\begin{abstract}
We present an analysis with improved sensitivity to
the light charged Higgs ($m_{H^+} < m_t-m_b$) searches in the top
quark decays $t \to b H^+ \to b (\tau^+\nu_\tau) + ~{\rm c.c.}$ in the $t\bar{t}$
and single $t/\bar{t}$ production processes at the LHC.
In the Minimal Supersymmetric Standard Model (MSSM), one anticipates the
branching ratio ${\cal B} (H^+ \to \tau^+\nu_\tau)\simeq 1$ over almost the entire
allowed $\tan \beta $ range. Noting that
the $\tau^+$ arising from the decay $H^+ \to \tau^+\nu_\tau$ are predominantly right-polarized,
as opposed to the $\tau^+$ from the dominant background $W^+ \to \tau^+\nu_\tau$, which are
left-polarized, a number of $H^+/W^+ \to \tau^+\nu_\tau$ discriminators have been proposed
and studied in the literature. We consider hadronic decays of the $\tau^\pm$, concentrating on the
dominant one-prong decay channel $\tau^\pm \to \rho^\pm \nu_\tau$. The energy and $p_T$ of the charged
prongs normalised to the corresponding quantities of the $\rho^\pm$ are convenient variables 
which serve as $\tau^\pm$ polariser. We use the distributions in these variables and 
several other kinematic quantities  
to train a boosted decision tree (BDT). Using the BDT classifier, and a variant of it called
 BDTD, which makes use of decorrelated variables, we have calculated the BDT(D)-response functions
to estimate the signal efficiency vs. the rejection of the background. We argue that this chain of
analysis has a high sensitivity to light charged Higgs searches up to a mass of 150 GeV 
 in the decays $t \to  b H^+$ (and charge conjugate) at the LHC. 
 For the case of single top production, we
also study the transverse mass of the system determined using Lagrange multipliers.

 \end{abstract}

\maketitle

\section{Introduction}

In many extensions of the standard model (SM), the Higgs sector of the SM is enlarged by
adding an extra doublet of complex Higgs fields. After spontaneous symmetry breaking, one
finds three neutral Higgs bosons $(h,H,A)$ and a pair of charged Higgs bosons, $H^\pm$.
These neutral and charged Higgs bosons have been searched for in high energy experiments,
in particular, at LEP and the Tevatron. None of these Higgses have been seen so far, and
upper limits exist on all of them~\cite{Nakamura:2010zzi}.
We will concentrate here on the charged Higgs searches, in which the two key phenomenological
parameters are the charged Higgs mass, $m_{H^\pm}$, and $\tan \beta$, the ratio of the two
vacuum expectation values, $\tan \beta =v_2/v_1$. The searches for the $H^\pm$ are model-dependent,
and the exclusion limits (expressed as a contour in the $m_{H^\pm}$ - $\tan\beta$ plane) have to be
taken together with the underlying model.
For example, in the so-called two-Higgs-doublet-models (2HDM), a stringent limit exists on $m_{H^\pm}$
from the measured branching ratio for $B \to X_s \gamma$ and the NNLO estimates of the same in the
SM, yielding $m_{H^\pm} > 295 (230)$ GeV at the 95\% (99\%) C.L., for almost the entire $\tan \beta$
values of interest~\cite{Misiak:2006zs}. This limit can be easily evaded in other models, in particular, in the
minimal supersymmetric model (MSSM).  
 
 Direct $H^\pm$-searches
are limited by the center-of-mass energy in $e^+e^- \to H^+ H^-$ annihilation processes, where
they can be produced via $s$-channel exchange of a photon or a $Z$ boson.
These searches assume
for the branching ratios ${\cal B}(H^+ \to \tau^+ \nu_\tau)
+ {\cal B}(H^+ \to c\bar{s})=1$ and hold for all values of ${\cal B}(H^+ \to \tau^+ \nu_\tau)$.
 In the 2HDM framework,
the cross section in the Born approximation depends only on $m_{H^\pm}$ (modulo the known couplings)
 and the present limit is $m_{H^\pm} > 79.3$ GeV at 95\% C.L. obtained at
$E_{\rm cm}(e^+e^-)=209$ GeV from LEP~\cite{Nakamura:2010zzi}. The mass range $m_{H^+} < m_t-m_b$ has
been searched in the process $p\bar{p} \to t \bar{t} X$ at the Tevatron, followed by the
decay $t \to b H^+$ (and its charge conjugate).  For
example, Altonen {\it et al.}~\cite{Aaltonen:2009ke} have searched for the decay $t \to b H^+$,
 followed by
$H^+ \to c\bar{s}$ in 2.2 fb$^{-1}$ of $p\bar{p}$ collisions at $E_{\rm cm}(p\bar{p})=1.96$ TeV,
obtaining upper limits on ${\cal B}(t \to bH^+)$ between 0.08 and 0.32 (95\% C.L.), assuming
 ${\cal B}(H^+ \to c\bar{s})=1$. In the MSSM, this probes only a very small $\tan \beta$ region,
namely $\beta < 1$, which is not favoured by theoretical considerations~\cite{Carena:2110pdg}.
The search for $t \to b H^+$, followed by $H^+ \to \tau^+ \nu_\tau$ by Abazov
 {\it et al.}~\cite{Abazov:2009wy}
in 0.9$^{-1}$ of $p\bar{p}$ collisions at the Tevatron yield upper limits on ${\cal B}(t \to bH^+)$
 between 0.19 and
0.25 (95\% C.L.) for $m_{H^+}=80 - 155$ GeV and ${\cal B}(H^+ \to \tau^+ \nu_\tau)=1$.  This excludes
a small region ($\tan \beta > 35$ and $m_{H^+}=100 - 120$ GeV)~\cite{Stal:2010ay}. Thus, it is fair to conclude that
the searches of the charged Higgses over a good part of the $m_{H^\pm}$ - $\tan\beta$ plane in the
 MSSM is a programme that still has to be carried out and this belongs to the
LHC experiments. In anticipation, searches for the $H^\pm$ in $pp$ collisions at $E_{\rm cm}=7 - 14$ TeV
at the LHC have received a lot of
 attention~\cite{Aad:2009wy,ATLAS-2010-006,ATLAS-2010-003,Ball:2007zza,Baarmand:2006dm,Kinnunen:2008zz}.    
There are two regions, namely  $m_{H^+} < m_t-m_b$,  which will
be looked into in both the $t\bar{t}$ pair production and in single top (or anti-top) production
in $pp$ collisions, followed by the decays $t \to b H^+$ and  $H^+ \to \tau^+ \nu_\tau$, and for 
$m_{H^\pm}$ above the  top quark mass, in which case $H^\pm$ production mainly takes place through the
process $g b \to tH^+$, followed dominantly by the decay $H^+ \to t \bar{b}$. However, despite
larger branching fraction, it may be hard to distinguish the $H^+ \to t\bar{b}$ mode from the
bckground. For large $\tan \beta$, the decay mode $H^+ \to \tau^+ \nu_\tau$ becomes discernible. 
In this paper, we will concentrate on the
light $H^\pm$-scenario.

The decay channel $H^\pm \to \tau^\pm + \nu_\tau$ will play the key role in the searches
of the light $H^\pm$-bosons. The $\tau^+$ leptons arising
from the decays $W^+ \to \tau^+ \nu_\tau$ and $H^+\to \tau^+ \nu_\tau$ are predominantly
left- and right-polarised, respectively. Polarisation of the $\tau^\pm$  influences the
 energy distributions in the subsequent decays of the $\tau^\pm$.
 Strategies to enhance the $H^\pm$-induced effects in the decay
 $t \to b (W^+,H^+) \to b (\tau^+ \nu_\tau)$,  based on the
polarisation of the $\tau^+$ have been discussed at length, starting from the pioneering
 work~\cite{Hagiwara:1989fn,Bullock:1991fd,Rouge:1990kv,Bullock:1992yt} to the production and
decays of a $t\bar{t}$ pair at the hadron colliders Tevatron and the
 LHC~\cite{Roy:1991sf,Raychaudhuri:1995kv,Roy:1999xw,Assamagan:2002in,Cao:2003tr}.
 Also the effects of the (QED and QCD) radiative corrections on such distributions in
the dominant (one-charged prong) decay channels
 $ \tau^+ \to \pi^+ \nu_\tau, \rho^+ \nu_\tau, a_1^+ \nu_\tau$
and $\ell^+ \bar{\nu}_\ell \nu_\tau $ have been worked out~\cite{Ali:2009sm}.
Following these studies, the construction of the $\tau^\pm$-jet (as well as $b$-jet) are of central
importance in $H^\pm$-searches. We use the dominant single-charged-prong decay
 $\tau^\pm \to \rho^\pm \nu_\tau$ as the $\tau^\pm$ polariser.  As $\rho^\pm \to \pi^\pm \pi^0$ is
the dominant decay mode, the energy and transverse momentum of
 the $\pi^\pm$ in the $\tau^\pm$-jet become quantities of main interest for our study. 
Likewise, the distribution in the angle $\psi$, defined as 
\begin{equation}
\cos \psi = \frac{2m_{\rho b}^2}{m_{\rm top}^2 - m_W^2} -1~,
\label{eq:angle-psi}
\end{equation}
plays an important role in our analysis.
 Since the energy-momentum vectors of
the $b$-jet and the $\rho^\pm$ can be measured, this distribution is measurable at the LHC. We also note
that this distribution is different from the conventional definition of the angle
 $\psi$~\cite{Eriksson:2007fx}, in which
the invariant mass $m_{\ell b}^2$ is measured instead of $m_{\rho b}^2 $. The other distributions
that enter in our analysis are listed in the next section.

Having generated these distributions, characterising the signal
 $t \to b H^+ \to b (\tau^+ \nu_\tau) \to  b (\rho^+ \bar{\nu}_\tau)\nu_\tau)$
and the background $t \to b W^+ \to b (\tau^+ \nu_\tau) \to  b (\rho^+ \bar{\nu}_\tau)\nu_\tau)$
 events, we use a technique called the
Boosted Decision Tree (BDT) -- a classification model used widely in data mining~\cite{Han:2006} 
-- to develop an identifier optimised for the $t \to b H^+$ decays. In our calculation, we use both
 BDT and a variant of it called BDTD (here D stands for decorrelated), where possible correlations
in the input variables are removed by a proper rotation obtained from the decomposition 
of the square root of the covariance matrix, to
discriminate the signal events from the large backgrounds.  We recall that
this technique has been successfully used to establish the single top quark production in $p \bar{p}$
collisions at the Tevatron~\cite{Abazov:2009ii,Aaltonen:2009jj} (see~\cite{Liu:2009zz} for details).
Recently, we have applied this technique to a feasibility study of measuring the CKM matrix element
$\vert V_{ts}\vert$ from the decay $t \to W s$ at the LHC@14 TeV, and have estimated that a benchmark with
10\% accuracy for this decay mode with a $10^3$ rejection of the background $t \to Wb$  can be achieved 
with an integrated luminosity of 10 (fb)$^{-1}$~\cite{Ali:2010xx}. We show in this paper that a similar
BDTD-based analysis holds great promise in light-$H^\pm$ searches at the LHC both in the 
$pp \to t\bar{t} X$
pair production and in the single top (or anti-top) production $pp \to t/\bar{t} X$. Furthermore,
we show that using a transverse mass definition, as suggested in~\cite{Gross:2009wia}, 
the process $pp \to t/\bar{t} X$ followed by the decays $t \to b H^+, bW^+$, allows one to
determine rather sharp Jacobian peaks for the transverse mass of the $H^\pm$-bosons. The conventional
definition of the transverse mass~\cite{Smith:1983aa}, which was very helpful in the determination
of the transverse mass of the $W^\pm$ bosons, is less suited for constructing the corresponding
mass of the $H^\pm$ bosons.

We note that an analysis using an iterative discriminat analysis method similar to the one
presented here was carried out  by Hesselbach {\it et al.}~\cite{Hesselbach:2008zz}.
In particular, detailed Monte Carlo comparisons of
several variables incorporating the spin effects in charged Higgs boson
production were presented to separate the $tbH^+$ signal from the standard
model  $t\bar{t}$ background both at the Tevatron ($\sqrt{s}=1.96$ TeV) and the
LHC ($\sqrt{s}=14$ TeV). However, there
are several significant  differences in the two studies, such as
the distribution in $\cos \psi$ (defined in eq. (1)), which plays
an important role in our analysis. In addition, we have studied the case of
single top production at the LHC, $pp \to t/\bar{t} +X$,
followed by the decays $t \to b( H^+/W^+ \to \tau^+\nu_\tau)$ +c.c., which was not considered in
Ref.~\cite{Hesselbach:2008zz}.

This paper is organised as follows: In section 2, we analyse the process $pp \to t\bar{t} X$ at the LHC,
followed by the decay chains $t \to bW^+, bH^+$, and the subsequent decays
$(H^+,W^+) \to \tau^+ \nu_\tau$, together with the BDTD-based analysis of the signal ($t \to bH^+$)
and the SM decay background ($t \to bW^+$). The BDTD response functions are then used
to work out the signal efficiency vs.~the background rejection. In section 3, we repeat this
analysis for the single top (or anti-top) production $pp \to t/\bar{t} X$ at the LHC. Section 4
contains a brief summary.

\section{$t\bar{t}$ production and the decay chains $t \to b W^+/H^+ \to b (\tau^+ \nu_\tau)$
at the LHC}

\subsection{ Production cross sections}
Theoretical predictions of the top quark production at the LHC have been
obtained by including up to the next-to-next-to-leading order (NNLO) corrections in the
strong coupling
 constant~\cite{Bonciani:1998vc,Cacciari:2008zb,Kidonakis:2008mu,Moch:2008qy}
using  modern parton distribution
functions (PDFs)~\cite{Martin:2007bv,Nadolsky:2008zw}.  Typical estimates for 
$\sigma(pp \to t\bar{t}X)$ range from  $874^{+14}_{-33}$ pb for
$m_t=173$ GeV and $\sqrt{s}=14$ TeV~\cite{Langenfeld:2009tc} to
 $943 \pm 4({\rm kinematics}) ^{+77}_{-49}({\rm scale})\pm 12 ({\rm PDF})$ pb~\cite{Kidonakis:2008mu}.
Compared to the $t\bar{t}$ production cross section at the Tevatron, this 
is larger by two orders of magnitude. The cross sections at the lower
LHC energies, 7 and 10 TeV, have also been calculated~\cite{Langenfeld:2009tc,Kidonakis:2008mu}, with
$\sigma (pp \to t\bar{t}X) \simeq 400$ pb at 10 TeV and about half that number
at 7 TeV. Thus, for the top quark physics, the dividends in going from 7 to 14 TeV
are higher by a good factor 4. 
\subsection{Top quark decays $t \to b (W^+, H^+)$ and charged Higgs decays $H^+ \to c\bar{s}, \tau^+\nu_\tau$}
\begin{figure}[ht]
\centering
\hspace*{-2cm}\includegraphics[width=12.5cm,height=9.5cm]{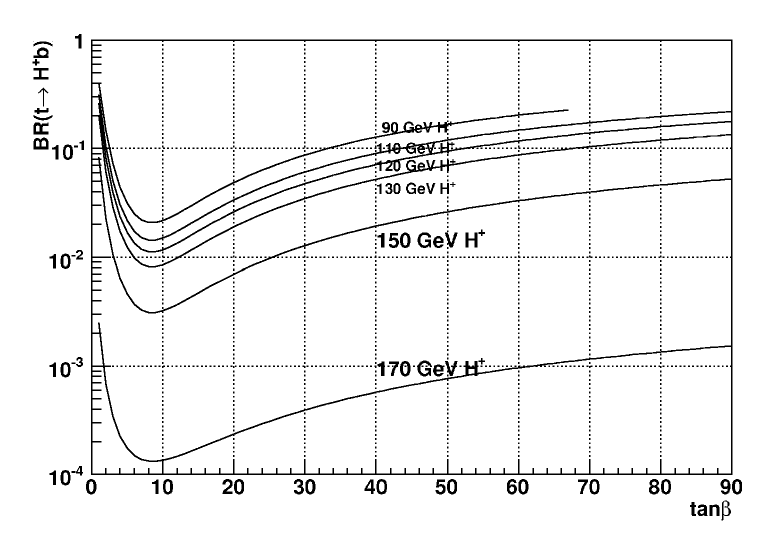}
\caption{\label{fig:fig1b-higgs.pdf}
Branching ratio ${\cal B}(t \to H^+b)$ in MSSM as a function of $\tan \beta$ for the charged
Higgs masses as indicated. (Figure taken from \cite{Sopczak:2009sm}).}
\end{figure}

Top-quark decays within the Standard Model are completely dominated by the
mode
\ba
    t \to b + W^+~, \label{ProcessW}
\ea
due to $V_{tb}=1$ to a very high accuracy.  In beyond-the-SM theories with an extended Higgs
sector, a light charged Higgs can also be produced via
\ba
    t \to b + H^+~. \label{ProcessH}
\ea
The relevant part of the interaction Lagrangian is \cite{Raychaudhuri:1995kv}:
\ba
    {\cal L}_I &=&
    \frac{g}{2\sqrt{2} M_W} V_{tb}
    H^+ \brs{\bar u_t\br{p_t} \brf{A\br{1+\gamma_5}+B\br{1-\gamma_5}} u_b\br{p_b}}
    \nn\\&+&
    \frac{g C}{2\sqrt{2} M_W}
    H^+ \brs{\bar u_{\nu_l}\br{p_\nu} \br{1-\gamma_5} u_l\br{p_l}},
    \label{HiggsLagrangian}
\ea
where $A$, $B$ and $C$ are model-dependent parameters which
depend on the fermion masses and $\tan \beta$:
\ba
    A = m_t \cot\beta, \qquad B = m_b \tan \beta, \qquad C = m_\tau \tan \beta.
    \label{LagrangianParametersABC}
\ea
The decay widths of processes (\ref{ProcessW}) and (\ref{ProcessH}) in the Born approximation
are \cite{Raychaudhuri:1995kv}:
\ba
    \Gamma^{\rm Born}_{t \to b W} &=&
    \frac{g^2}{64\pi M_W^2 m_t}
    \lambda^{\frac{1}{2}}\br{1,\frac{m_b^2}{m_t^2},\frac{M_W^2}{m_t^2}}
    \brs{
        M_W^2\br{m_t^2+m_b^2} + \br{m_t^2-m_b^2}^2 - 2M_W^4
    },
    \\
    \Gamma^{\rm Born}_{t \to b H} &=&
    \frac{g^2}{64\pi M_W^2 m_t}
    \lambda^{\frac{1}{2}}\br{1,\frac{m_b^2}{m_t^2},\frac{M_H^2}{m_t^2}}
    \nn\\
    &\times&
    \brs{
        \br{m_t^2\cot^2\beta + m_b^2\tan^2\beta}
        \br{m_t^2 + m_b^2 - M_H^2}
        -
        4 m_t^2 m_b^2
    },
\ea
where $\lambda\br{x,y,z} = x^2+y^2+z^2-2xy-2xz-2yz$ is the
triangle function. The total top quark decay width in the Born approximation is obtained
by adding the two partial widths
\ba
    \Gamma_t^{\rm tot,~Born} = \Gamma^{\rm Born}_{t \to b W} + \Gamma^{\rm Born}_{t \to b H}.
    \label{TotalWidth}
\ea
QED corrections in the total decay width of the top quark are numerically small.
 The $O(\alpha_s)$ QCD corrections were calculated in
\cite{Czarnecki:1992ig,Czarnecki:1992zm} (see, also Ref.~\cite{Li:1990cp}) and have the form:
\ba
    \Gamma_{t,RC}^{\rm tot} &=& \Gamma_{t \to b W}^{\rm Born+QCD} + \Gamma_{t \to b H}^{\rm Born+QCD},
    \label{TotalWidthWithRC}
    \\
    \Gamma_{t \to b (W,H)}^{\rm Born+QCD} &=&
    \Gamma^{\rm tot, Born}_t\br{1 + f_{W,H}},
    \qquad
    f_{W,H} =
    \frac{\alpha_s}{3\pi}\br{5-\frac{4\pi^2}{3}}. \nn
\ea
Thus, in the branching ratio ${\cal B}(t\to bH^+)$, also this QCD correction drops out. 
However, radiative corrections coming from the supersymmetric sector to 
${\cal B}(t\to bH^+)$ are rather important. They have been calculated in great detail in the
literature, in particular for the MSSM scenario in \cite{Guasch:1995rn,Coarasa:1996qa,Carena:1999py}, and can be
effectively incorporated
by replacing the $b$-quark mass $m_b$ in the Lagrangian for the decay
 $t \to  b H^+$ by the SUSY-corrected mass
 $m_b^{\rm corrected}=m_b/[1 + \Delta_b]$. The correction $\Delta_b$ is a
 function of the supersymmetric parameters and, for given MSSM scenarios,
this can be calculated using the FeynHiggs programme~\cite{Hahn:2008zzd}.
In particular, for large values of $\tan \beta$ (say, $\tan \beta > 20)$),
the MSSM corrections increase the branching ratio for $t \to bH^+$ significantly.
This, for example, can be seen in a
particular MSSM scenario in a recent update~\cite{Sopczak:2009sm}, from where we show
${\cal B}(t\to bH^+)$ as a function of $\tan \beta$, calculated for $m_t=175$ GeV and various assumed values
of the charged Higgs mass, indicated in Fig.~\ref{fig:fig1b-higgs.pdf}. 

Since we are treating the case of the light charged Higgs, there are essentially only two
decay modes which are important: $H^+ \to \tau^+\nu_\tau$ and $H^+ \to c \bar{s}$.
The branching ratio of interest to us ${\cal B} (H^+ \to \tau^+\nu_\tau)$ is given by
\cite{Raychaudhuri:1995kv}:
\ba
 {\cal B} (H^+ \to \tau^+\nu_\tau)   &=& \frac{\Gamma_{H \to \tau \nu_\tau}}{\Gamma_{H \to \tau \nu_\tau}+\Gamma_{H \to c\bar s}}~, \\
    \Gamma_{H \to \tau \nu_\tau} &=&
    \frac{g^2 M_H}{32 \pi M_W^2} m_\tau^2 \tan^2 \beta, \nn\\
    \Gamma_{H \to c\bar s} &=&
    \frac{3 g^2 M_H}{32 \pi M_W^2} \br{m_c^2 \cot^2 \beta + m_s^2 \tan^2\beta}. \nn
\ea
For the numerical values of $\tan \beta$ that we entertain in this paper,
the branching ratio  ${\cal B} (H^+ \to \tau^+\nu_\tau) =1$, to a very high accuracy.

\subsection{Event generation, trigger}
\begin{figure}[t]
\includegraphics[width=6.5cm,height=6.5cm]{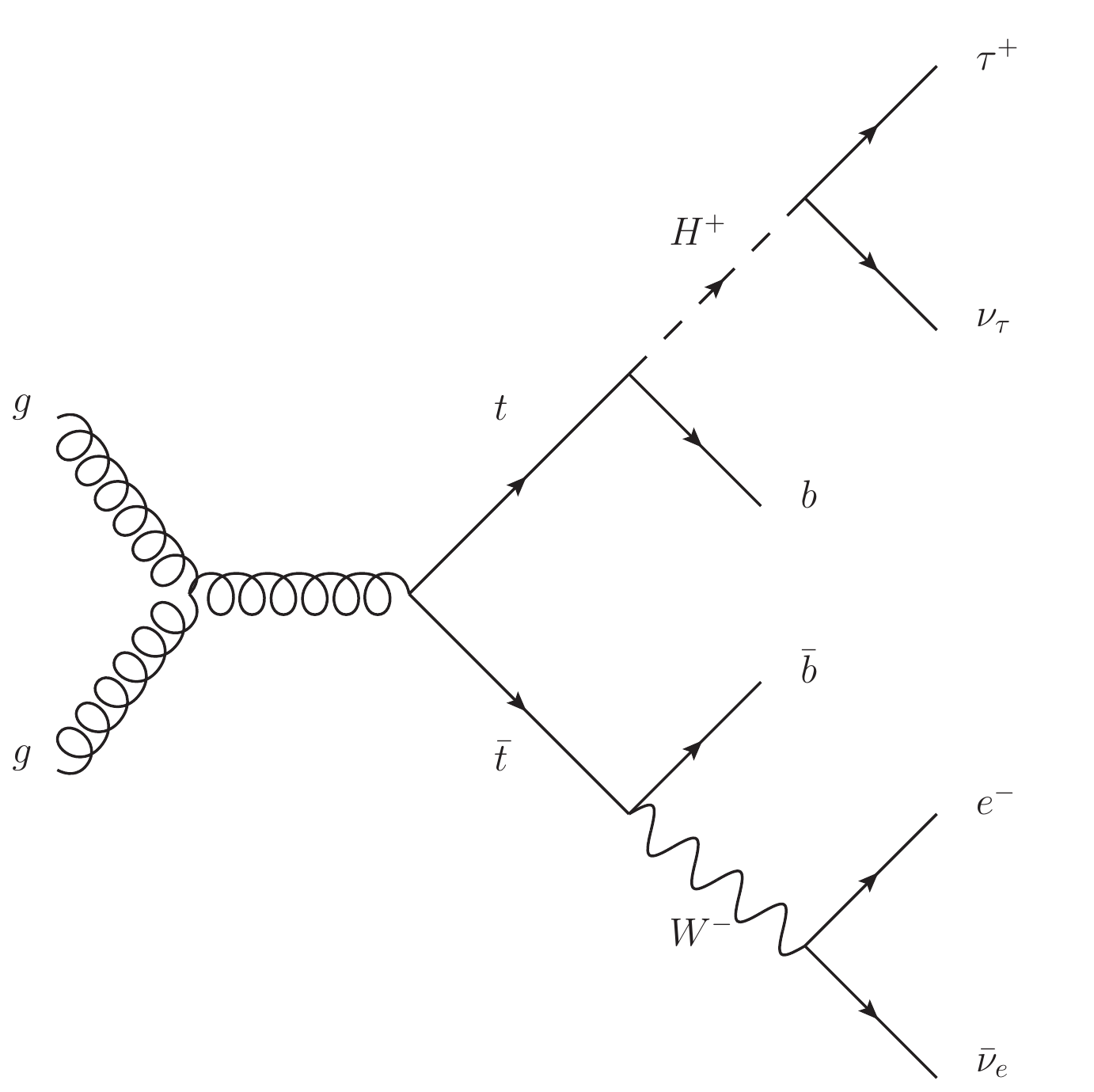}
\caption{\label{fig:feyn-ttbar}
Feynman diagram for $gg \to t\bar{t}$, followed by the decay $t \to b (H^+ \to \tau^+ \nu_\tau)$
and $\bar{t} \to \bar{b} (W^- \to e^- \bar{\nu}_e)$. }
\end{figure}
We consider in this section the process $pp \to t \bar{t}X$, with both the $t$ and $\bar{t}$
decaying into $Wb$. Our trigger is the leptonic decay $ W^- \to e^- \bar{\nu}_e$ or
 $ W^- \to \mu^- \bar{\nu}_\mu$.
The other $W^+$ decays via $W^+ \to \tau^+ \nu_\tau$. This makes up our main background. The signal
events are generated in which one of the $t$ or $\bar{t}$ decays via $W^+ \to bH^+$ (or its
charge conjugate $W^- \to bH^-$), see Fig.~\ref{fig:feyn-ttbar}. The other $\bar{t}$ or $t$ then decays leptonically, as in our trigger.
In the Minimal Supersymmetric Standard Model (MSSM), for large $\tan \beta$ and $m_{H^+} < m_t$,
the branching ratio for the decay $H^+ \to c\bar{s}$ is small and one anticipates the
branching ratio ${\cal B} (H^+ \to \tau^+\nu_\tau) \simeq 1$. This is the parameter space in which
the analysis reported here is valid.
  Noting that
the $\tau^+$ arising from the decay $H^+ \to \tau^+\nu_\tau$ are predominantly right-polarized,
as opposed to the $\tau^+$ from the dominant background $W^+ \to \tau^+\nu_\tau$, which are
left-polarized, a number of $H^+/W^+ \to \tau^+\nu_\tau$ discriminators have been proposed
and studied in the literature. We have used the dominant single-charged-prong decay $\tau^+ \to \rho^+ \nu_\tau$
as the $\tau^+$ polariser. Having set these branchings, we have generated 50K events for the process
$ p p \to t\bar{t} \to bW^+ (\bar{b}W^-)$, with all of them decaying according to the chain described earlier,
i.e., $W^- \to e^- \nu_e$ and $W^+ \to \tau^+ \nu_\tau$, with all the $\tau$'s forced to
decay into $\rho + \nu_\tau$ (here and below, charge conjugates are implied). In calculating the required
luminosity, we take into account the corresponding branching ratios, which are as
 follows~\cite{Nakamura:2010zzi}
\begin{eqnarray}
{\cal B}(W^+ \to e^+ \nu_e) &=& (10.75\pm 0.13)\%~,\nonumber\\
 ~~{\cal B}(W^+ \to \tau^+ \nu_\tau) &=& (11.25\pm 0.20)\%~,
\nonumber\\
{\cal B}(H^+ \to \tau^+ \nu_\tau)&=&1.0~,\nonumber\\
~~{\cal B}(\tau^+ \to \rho^+ \nu_\tau) &=&(25.5 \pm 0.10)\%~.
\end{eqnarray}  
We also generate the same number (50K) signal events, for each of the following charged Higgs masses:
$m_{H^+}=90, 110, 130, 150$ GeV. As for the background process, we force the $\tau^+$ to decay into
$\rho^+ \nu_\tau$ 100\% of the time. These events are generated using PYTHIA 6.4~\cite{Sjostrand:2006za}
and for the decays of the $\tau^\pm$, we use the programme called TAUOLA~\cite{Jadach:1993hs}
to incorporate the $\tau^\pm$ polarization information on the decay distributions.

We impose the following acceptance and trigger cuts:
\begin{itemize}
\item $|\eta_\ell| < 2.5$, with $\ell =e, \tau$
\item $|\eta_{b,\bar{b}}| < 2.5$
\item $P_{T_e} > 20$ GeV
\item $P_{T_\rho} > 10$ GeV
\item $P_{T_{b,\bar{b}}} > 20$ GeV
\end{itemize}

In order to discriminate the signal and background, we have studied a number of distributions, summarized
below.
\begin{itemize}
\item Distribution in the angle $\psi$, defined in eq.~\ref{eq:angle-psi}.
This is defined for both the decay chains:
 $ t \to b W \to b (\tau \nu_\tau) \to b (\rho \bar{\nu}_\tau) \nu_\tau$ and 
$ t \to b H \to b (\tau \nu_\tau) \to b (\rho \bar{\nu}_\tau) \nu_\tau$. Since the energy-momentum vectors of
the $b$-jet and the $\rho^\pm$ can be measured, this distribution is measurable at the LHC. We also note
that this distribution is different from the conventional definition of the angle
 $\psi$~\cite{Eriksson:2007fx}, in which
the invariant mass $m_{\ell b}^2$ is measured instead of $m_{\rho b}^2 $.

\item Energy and $p_T$ of the $b$-jets from the decays $t \to b W^+$ and $t \to b H^+$.

 \item Energy and $p_T$ of the $\tau^+$ jets from the decays $W^+ \to \tau^+ \nu_\tau$
and $H^+ \to \tau^+ \nu_\tau$, concentrating on the
single-charged-prong decays $\tau^+ \to \rho^+ \nu_\tau$.

\item The ratio of the energy and $p_T$ of the
$\tau^+$ jets and their accompanying $b$-jet.

\item As a measure of the $\tau$ polarisation, we consider the fractional energy  and transverse
momentum of the single-charged prong ($\pi^+$ in $\tau^+$-jet).

\item  For the case of single top production, we
also study the transverse mass of the system determined using Lagrange multipliers~\cite{Gross:2009wia}.

\item These
 distributions are
used to train a boosted decision tree (BDT). Using the BDT classifier, and a variant of it called
 BDTD, which makes use of decorrelated variables, we have calculated the BDT(D)-response functions
to estimate the signal efficiency vs. the rejection of the background.

\end{itemize}

The strategy adopted by us to search for the decays $t \to bH^+$ is somewhat different from the
traditional cut-based analysis, as, for example, reported in ~\cite{ATLAS-2010-006}. There
the idea is to suppress the SM-background as much as possible, making use of additional variables,
such as the missing $E_T$, satisfying $E_T^{\rm miss} > 50$ GeV. Our idea is, instead, to train
 a boosted decision tree classifier for both the signal and background events. Eventually, for
a realistic analysis of the LHC data, we may have to reintroduce some of the cuts to suppress
other non-$t\bar{t}$ background, such as coming from the process $pp \to W^\pm + {\rm jets}$,
which may also fake our signal.

\subsection{Details of the Analysis }
\begin{figure}[ht]
\centering
\includegraphics[width=0.99\textwidth,height=0.5\textwidth]{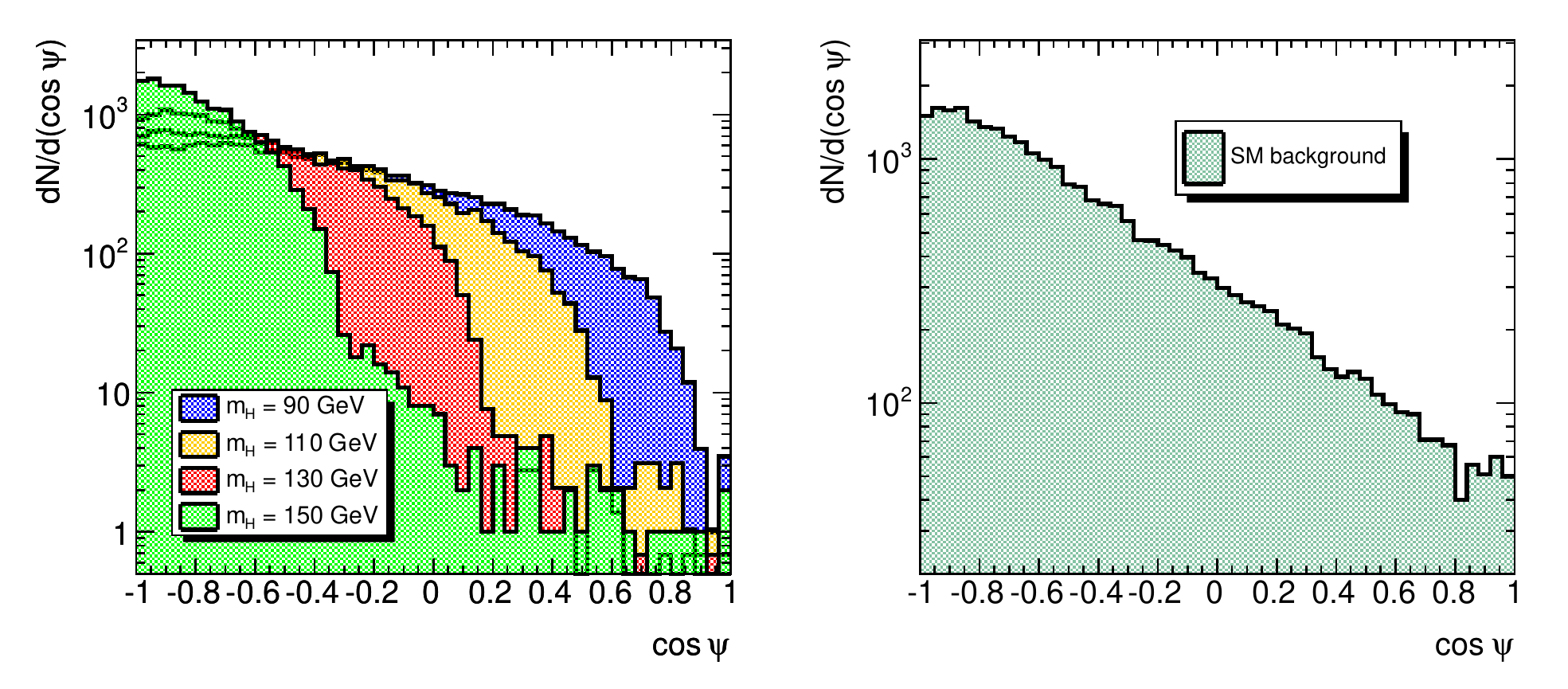}
\caption{\label{fig:cos-psi-ttbar}
The distribution $dN/d \cos \psi$ for the $t\bar{t}$ production as measured in the decay chain
$t \to b W \to b (\tau \nu_\tau) \to b (\rho \bar{\nu}_\tau) \nu_\tau)$ (right-hand frame),
and in $ t \to b H \to b (\tau \nu_\tau) \to b (\rho \bar{\nu}_\tau) \nu_\tau)$ for four
different charged Higgs masses, as indicated on the figure (left-hand frame).}
\end{figure}

In Fig.~\ref{fig:cos-psi-ttbar} (right-hand frame), we show the $\cos \psi$ distributions for the standard model (SM)
process $ p + p \to t\bar{t} +X$, followed by the decay chain 
$t \to b W \to b (\tau \nu_\tau) \to b (\rho \bar{\nu}_\tau) \nu_\tau)$. In the
left-hand frame, we show the same distribution when one of the $t$ or $\bar{t}$ decays via
the chain  $ t \to b H \to b (\tau \nu_\tau) \to b (\rho \bar{\nu}_\tau) \nu_\tau)$ , for four different
charged Higgs masses, as already stated in the previous section. For lower values of  $m_{H^+}$, the
$\cos \psi$ distribution falls less steeply than the SM background. 
 As $m_{H^+}$ increases, the  $\cos \psi$ distributions become steeper and are essentially confined
to the negative values of $\cos \psi$. This distribution then provides one of the discriminators to
be fed to the BDTD analysis.

\begin{figure}[ht]
\centering
\hspace*{-2cm}\includegraphics[width=14.5cm,height=10.5cm]{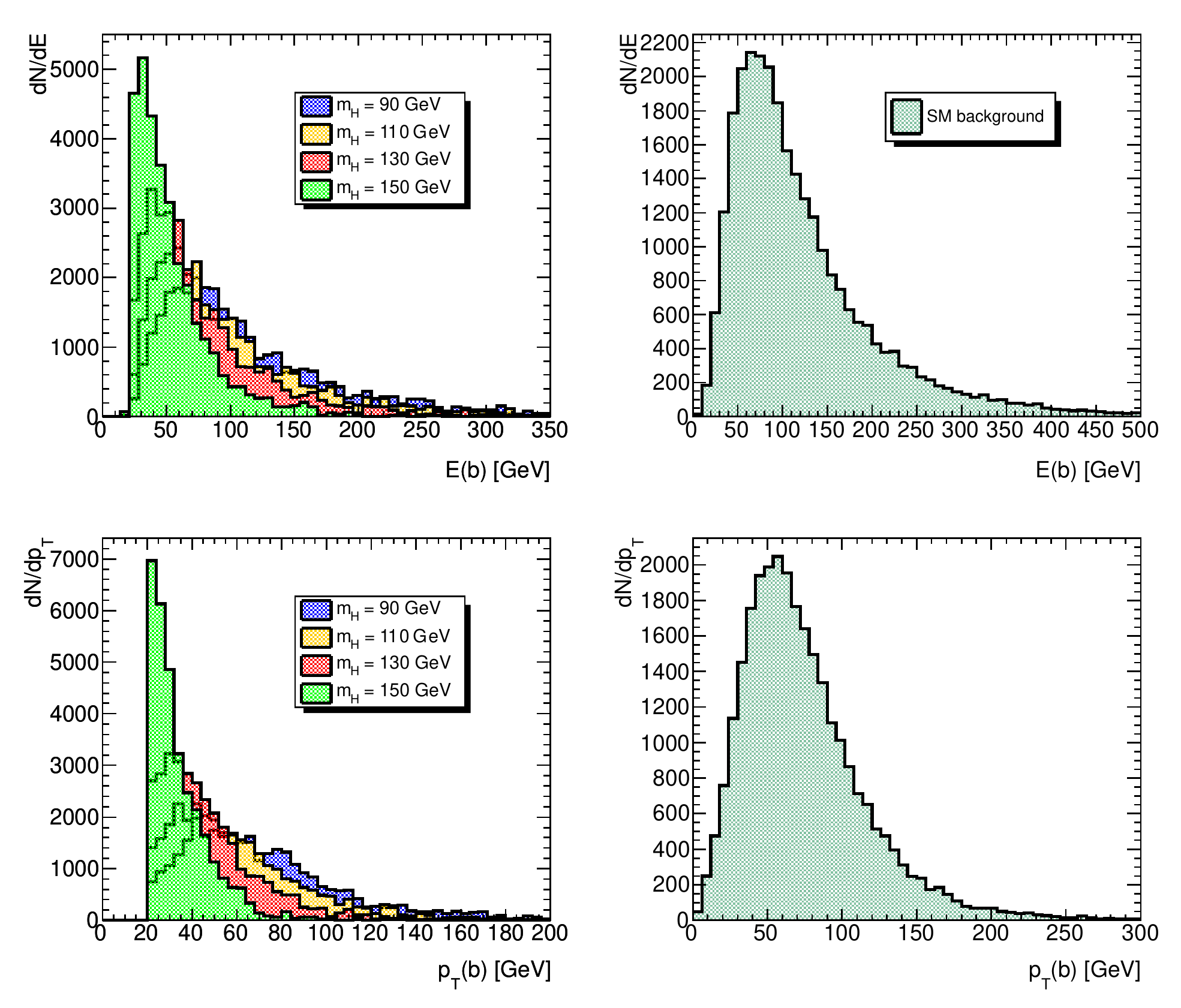}
\caption{\label{fig:bjets-ttbar}
Distributions in the Energy of the $b$-jet, $E(b)$, and transverse momentum of the $b$-jet, $p_T(b)$
 from the
process $p p \to t\bar{t} X$, followed by the decay $t \to W^+ b$ (right-hand frames), and
the same distributions for the decay chain $t \to H^+ b$ with the four indicated charged Higgs masses
(left-hand frame).}
\end{figure}

In Fig.~\ref{fig:bjets-ttbar} (right-hand frames), we show the distributions in the energy of the $b$-jet, $E(b)$,
 and the transverse momentum of the $b$-jet, $p_T(b)$ from the
 SM process process $p p \to t\bar{t} X$, followed by the decay chain discussed above. 
In the left-hand frames, the corresponding distributions are shown for the charged Higgs case. 
We remark that for the charged Higgs case these distributions are softer than those from the
SM due to the different helicity structure of the decays. This effect becomes stronger as 
$m_{H^+}$ increases due to phase space. As a result, these distributions add to the
discrimination power of the BDTD analysis. Note that these distributions reflect the
event characteristics at the generation level. Obviously, due to the semileptonic decays of the $b$-quark,
and other detector effects, they  will be modified. However, we expect that the dilutions
due to these effects are sub-dominant.

\begin{figure}[ht]
\centering
\hspace*{-2cm}\includegraphics[width=14.5cm,height=10.5cm]{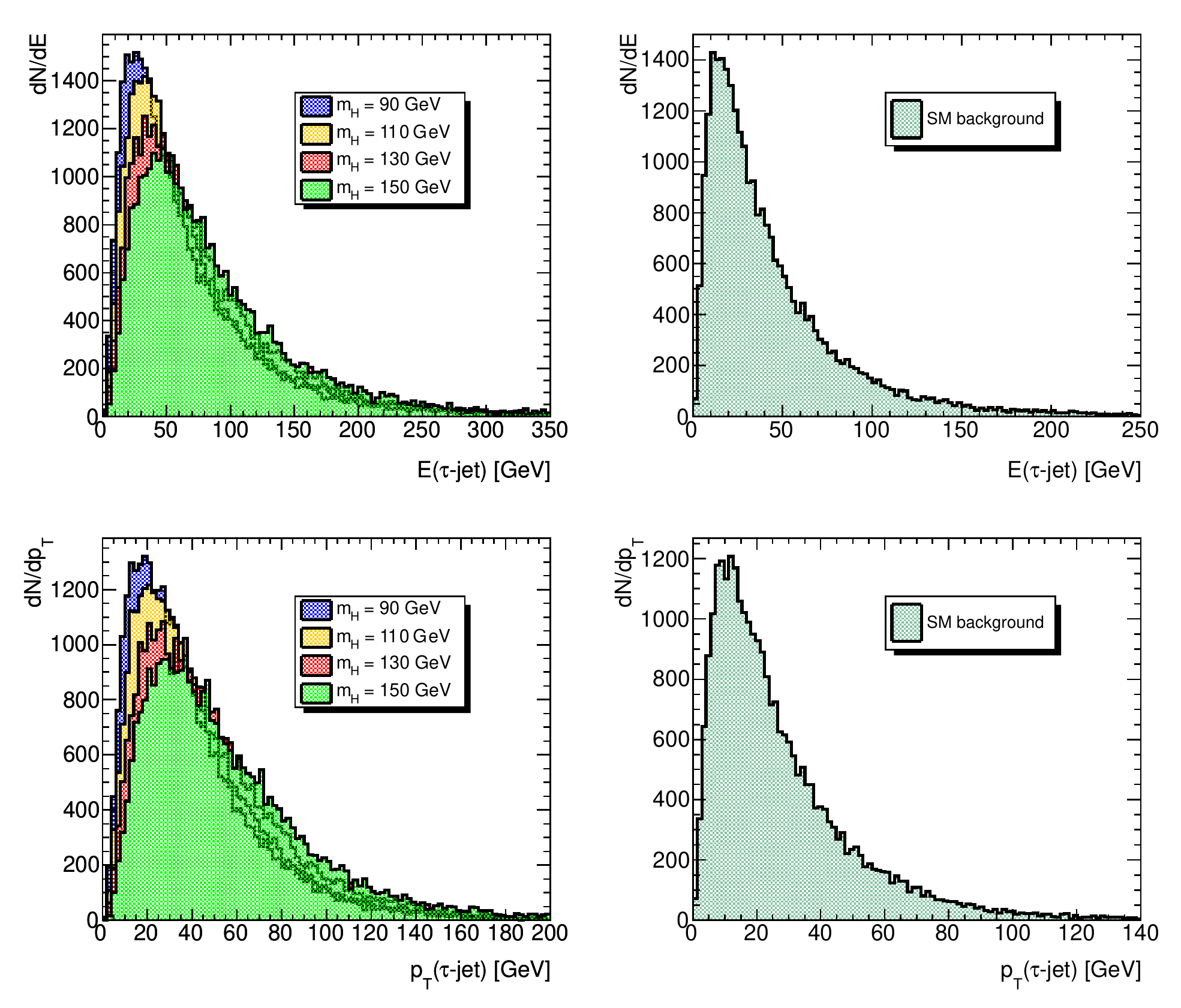}
\caption{\label{fig:tau-ttbar}
Distributions in the Energy of the $\tau$-jet, $E(\tau-{\rm jet})$, and transverse momentum of the
 $\tau$-jet, $p_T(\tau-{\rm jet}) $ from the
process $p p \to t\bar{t} X$, followed by the decay $t \to W^+ b$ (right-hand frames), and
the same distributions for the decay chain $t \to H^+ b$ with the four indicated charged Higgs masses
(left-hand frame).}
\end{figure}

In Fig.~\ref{fig:tau-ttbar} (right-hand frames), we show the distributions in the energy of the
 $\tau$-jet, $E(\tau-{\rm jet})$, and in the transverse momentum of the
 $\tau$-jet, $p_T(\tau-{\rm jet}) $ from the SM process, followed by the decay chain discussed above.
In the left-hand frames, the corresponding distributions are shown for the charged Higgs case.
In these distributions, the energy and $p_T$-spectra of the  $\tau$-jet coming from the charged
Higgs decays are harder than those coming from the SM process, and this difference becomes more marked
as $m_{H^+}$ increases. This complementary behaviour is expected for the same reason as discussed
earlier for Fig.~\ref{fig:bjets-ttbar}, again reflecting the differences in helicity and phase space.
It goes without saying that these distributions increase the discrimination power of the BDTD analysis.

\begin{figure}[ht]
\centering
\hspace*{-2cm}\includegraphics[width=14.5cm,height=10.5cm]{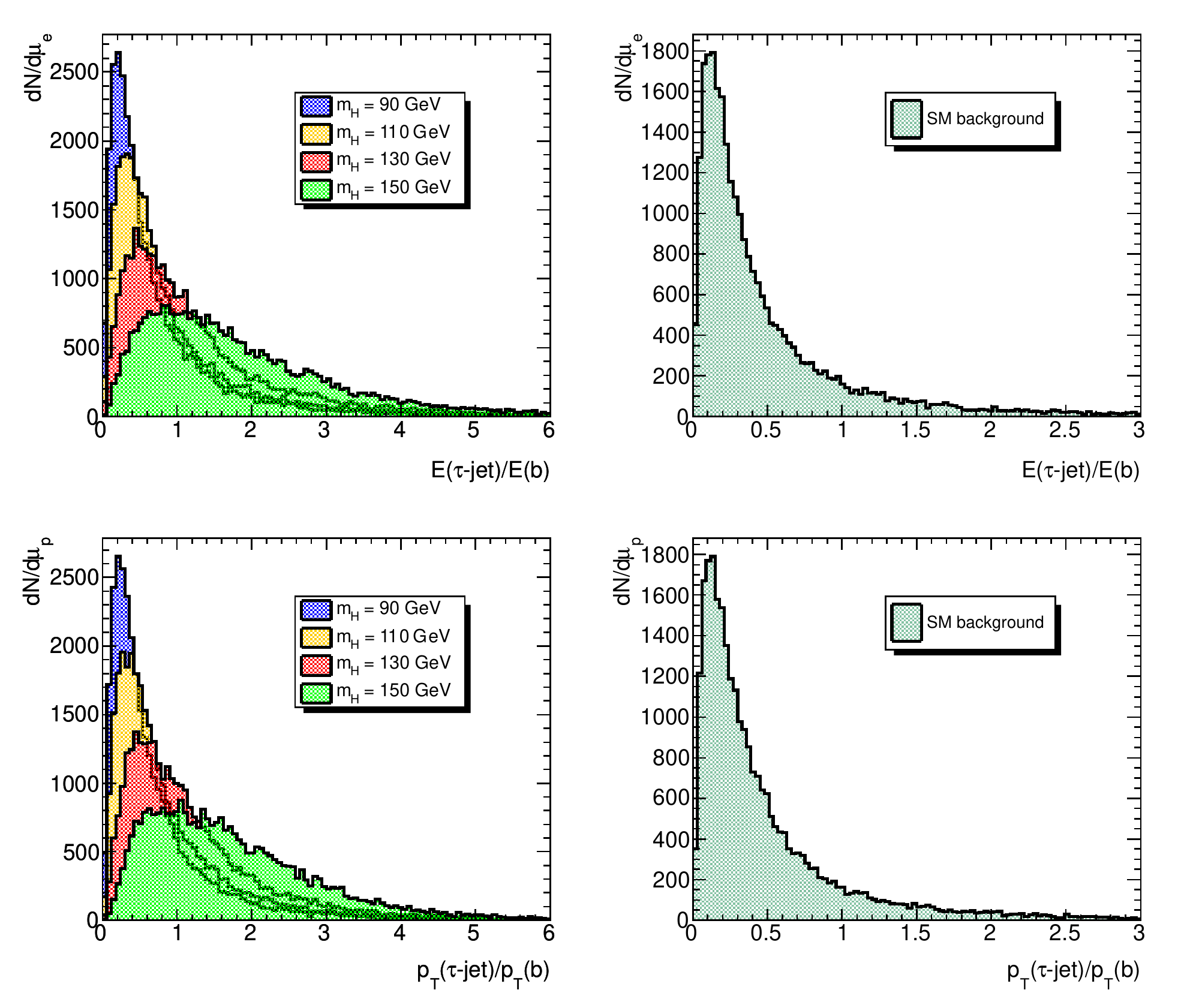}
\caption{\label{fig:tauB-ttbar}
Distributions in the ratio $E(\tau-{\rm jet})/E(b)$ and $p_T(\tau-{\rm jet})/p_T(b)$
from the
process $p p \to t\bar{t} X$, followed by the decay $t \to W^+ b$ (right-hand frames), and
the same distributions for the decay chain $t \to H^+ b$ with the four indicated charged Higgs masses
(left-hand frame).}
\end{figure}

To make this effect more marked, we show the ratio of the energy and $p_T$-spectra
involving the $\tau$- and $b$-jets in Fig.~\ref{fig:tauB-ttbar}. The SM distributions are shown in the
right-hand frames, and those from the charged Higgs in the left-hand frames. These distributions show
clearly the different shapes of the distributions SM vs. Higgs. For example, putting a lower cut
on the ratios $E(\tau-{\rm jet})/E(b) >1 $ or  $p_T(\tau-{\rm jet})/p_T(b) > 1$, most of the SM background
is eliminated, whereas the charged Higgs-induced distributions surviving this cut are considerably
larger, with the discrimination becoming stronger as $m_{H^+}$ increases.

\begin{figure}[ht]
\centering
\hspace*{-2cm}\includegraphics[width=14.5cm,height=10.5cm]{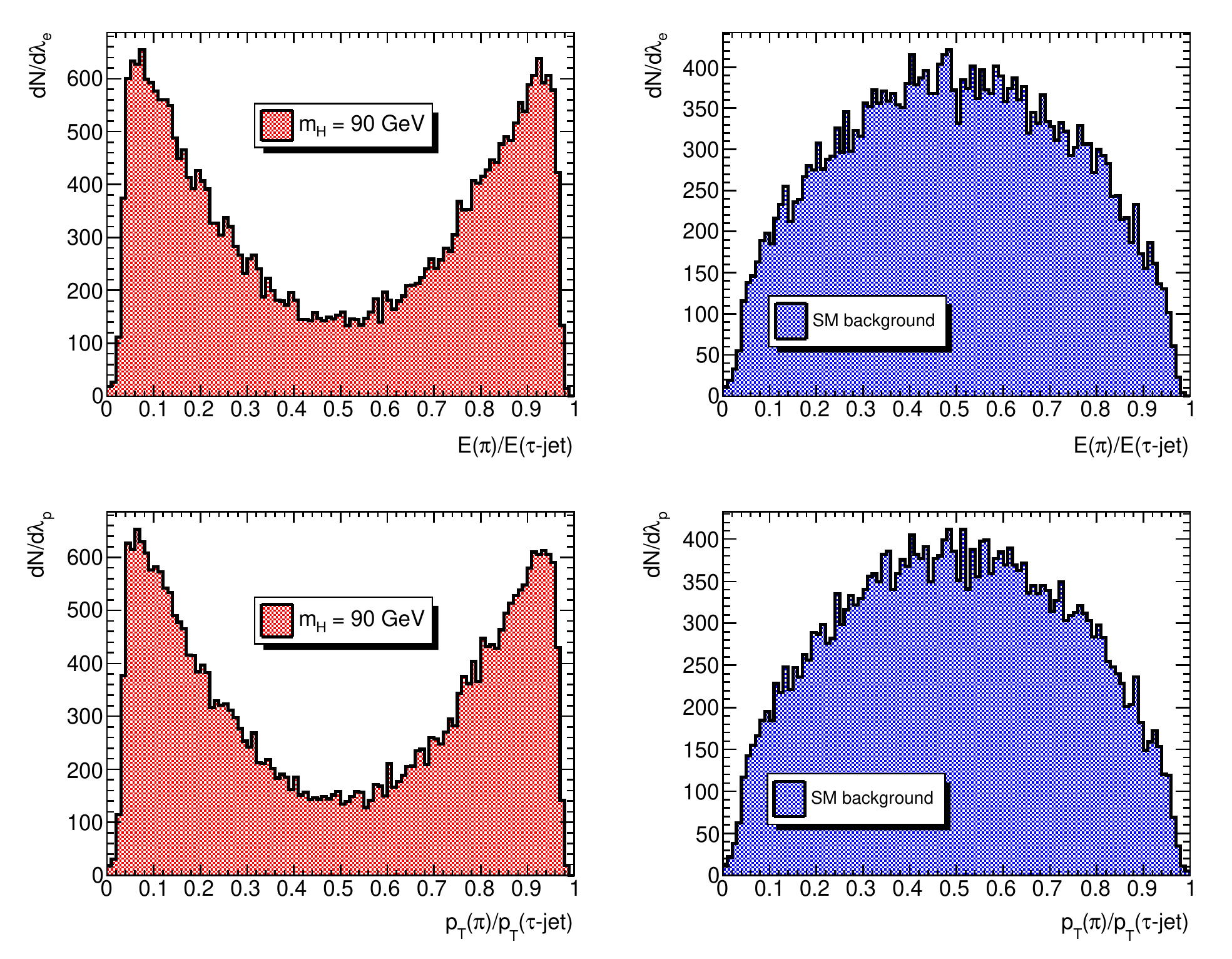}
\caption{\label{fig:polarization-ttbar}
Distributions in the fractional energy of the single-charged prong ($\pi^+$ in $\tau^+$-jet),
$E(\pi)/E(\tau-{\rm jet})$, and in the transverse
momentum of the single-charged prong, $p_T(\pi)/p_T(\tau-{\rm jet})$ from the
$p p \to t\bar{t} X$, followed by the decay $t \to W^+ b$ (right-hand frames), 
and
the same distributions for the decay chain $t \to H^+ b$ with the four indicated charged Higgs masses
(left-hand frame).}
\end{figure}

In Fig.~\ref{fig:polarization-ttbar}, we show the distributions in the fractional energy of the
 single-charged prong ($\pi^+$ in $\tau^+$-jet),
$E(\pi)/E(\tau-{\rm jet})$, and in the transverse
momentum of the single-charged prong, $p_T(\pi)/p_T(\tau-{\rm jet})$ from the SM process (right-hand frames)
and those coming from the charged Higgs-induced process (left-hand frames) for $m_{H^+}=90$ GeV.
 As remarked earlier. we are using the dominant
single-charged-prong decay $\tau^+ \to \rho^+\bar{\nu}_\tau$ as the $\tau^+$ polariser. 
As already noted in ~\cite{Bullock:1991fd},  the fractional energy distributions in $z=E_A/E_\tau$,
from the $\tau$-decay products $\tau \to A + $ missing energy, the effect of the
$\tau^\pm$ polarization  is most marked
for the decays $\tau^+\to \pi^+ \bar{\nu}_\tau$ and 
 $\tau^+\to \rho^+ \bar{\nu}_\tau$. This has been worked out in the
collinear limit, i.e., for $E_\tau/m_\tau \gg 1$. Our variables differ from the one used in
\cite{Bullock:1991fd}, in that we normalize to the visible $\tau$-energy and the visible $p_T(\tau-{\rm jet})$,
and not to the total $\tau$-energy. With our normalization, the $\pi^+$-energy measured in the
decays $\tau^+ \to \pi^+\bar{\nu}_\tau$ will be a delta function, peaked at 1 in the variables
shown in Fig.~\ref{fig:polarization-ttbar}, and hence we concentrate on  the 
decay chain $\tau^+\to \rho^+ \bar{\nu}_\tau$. These distributions also provide strong discriminants
for the BDTD analysis. 

\begin{figure}[ht]
\centering
\hspace*{-2cm}\includegraphics[width=14.5cm,height=10.5cm]{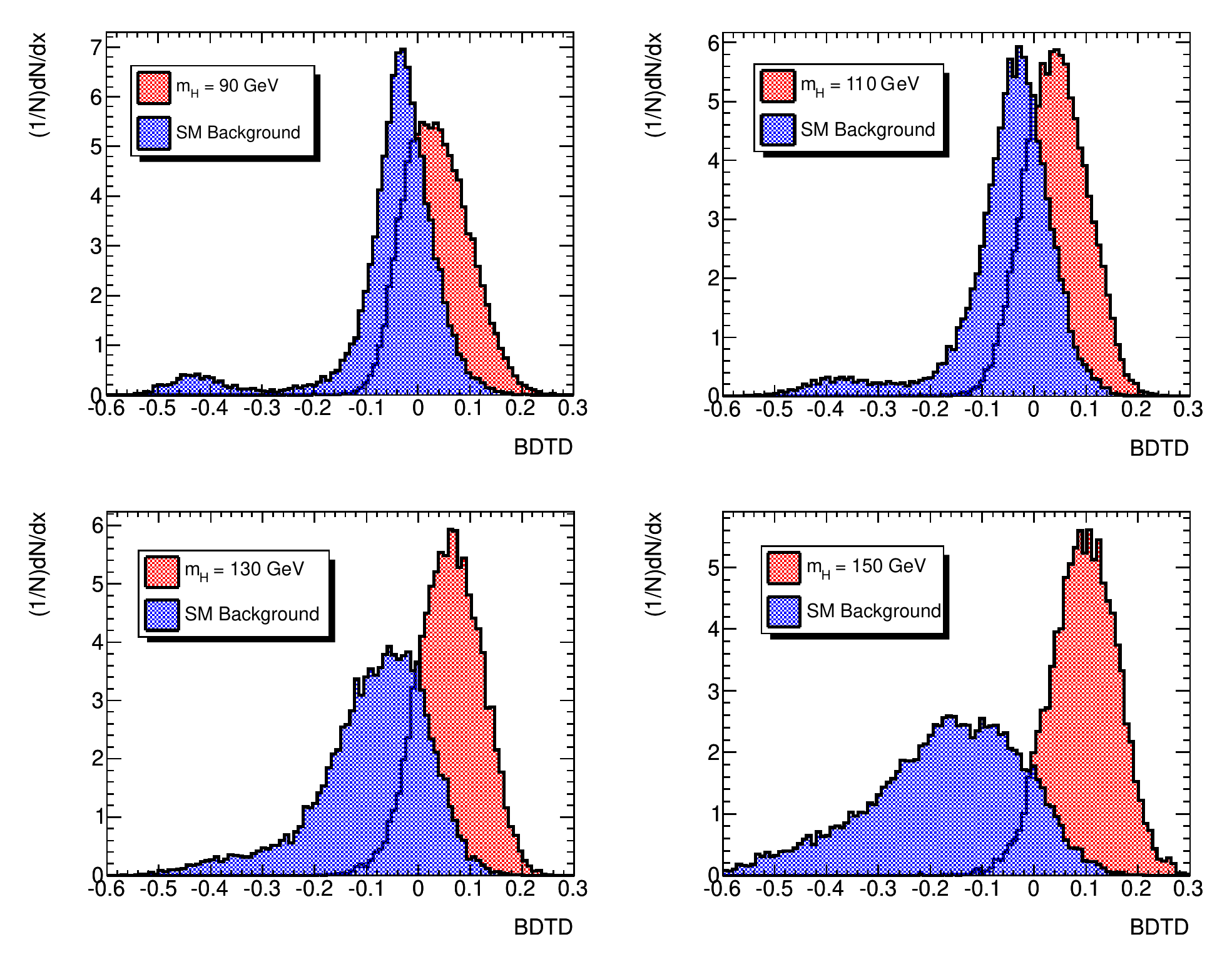}
\caption{\label{fig:bdtd-ttbar}
BDTD response functions for $p p \to t\bar{t} X$, with $\sqrt{s}=14$ TeV, followed by the decay
 $t \to W^+ b$ (SM) and the decay chain $t \to H^+ b$, with
 the SM background (in shaded blue) and the charged Higgs signal
process (in shaded red) for four different charged Higgs masses.}
\end{figure}

 Briefly, the generated input is used for the purpose of training and testing the samples.
 We provide the input in terms of the variables discussed earlier for the signal
($ t \to b H^+$) and the background ($ t \to b W^+$), obtained with the help of a Monte Carlo
generator. This information is used to develop the splitting criteria to determine the
best partitions of the data into signal and background to build up a decision tree (DT). 
The separation algorithm
used in splitting the group of events in building up DT plays an important role in the performance.
 The software 
called the Toolkit for Multivariate Data Analysis in ROOT (TMVA)~\cite{Hocker:2007ht} is used for the
 BDT(D) responses in our analysis. The response functions for $pp \to t\bar{t}X$ at a
center-of-mass-energy $\sqrt{s}=14$ TeV at the LHC, followed by the background process
$t \to b W^+$ (in shaded blue) and the signal  $t \to bH^+$ (in shaded red) are shown in
Fig.~\ref{fig:bdtd-ttbar}. The four frames shown in this figure correspond to the charged Higgs masses
$m_{H^+}=90, 110, 130$ and 150 GeV. As can be seen that the separation between the signal and the
background increases as $m_{H^+}$ increases. This improved separation as a function of $m_{H^+}$ will,
however, be compensated to some extent by the decreasing branching ratio for the decay
$t \to bH^+$, as shown in Fig.~\ref{fig:fig1b-higgs.pdf}~\cite{Sopczak:2009sm}, obtained by
using FeynHiggs~\cite{Heinemeyer:1998yj}.
 
\begin{figure}[ht]
\centering
\hspace*{-2cm}\includegraphics[width=14.5cm,height=10.5cm]{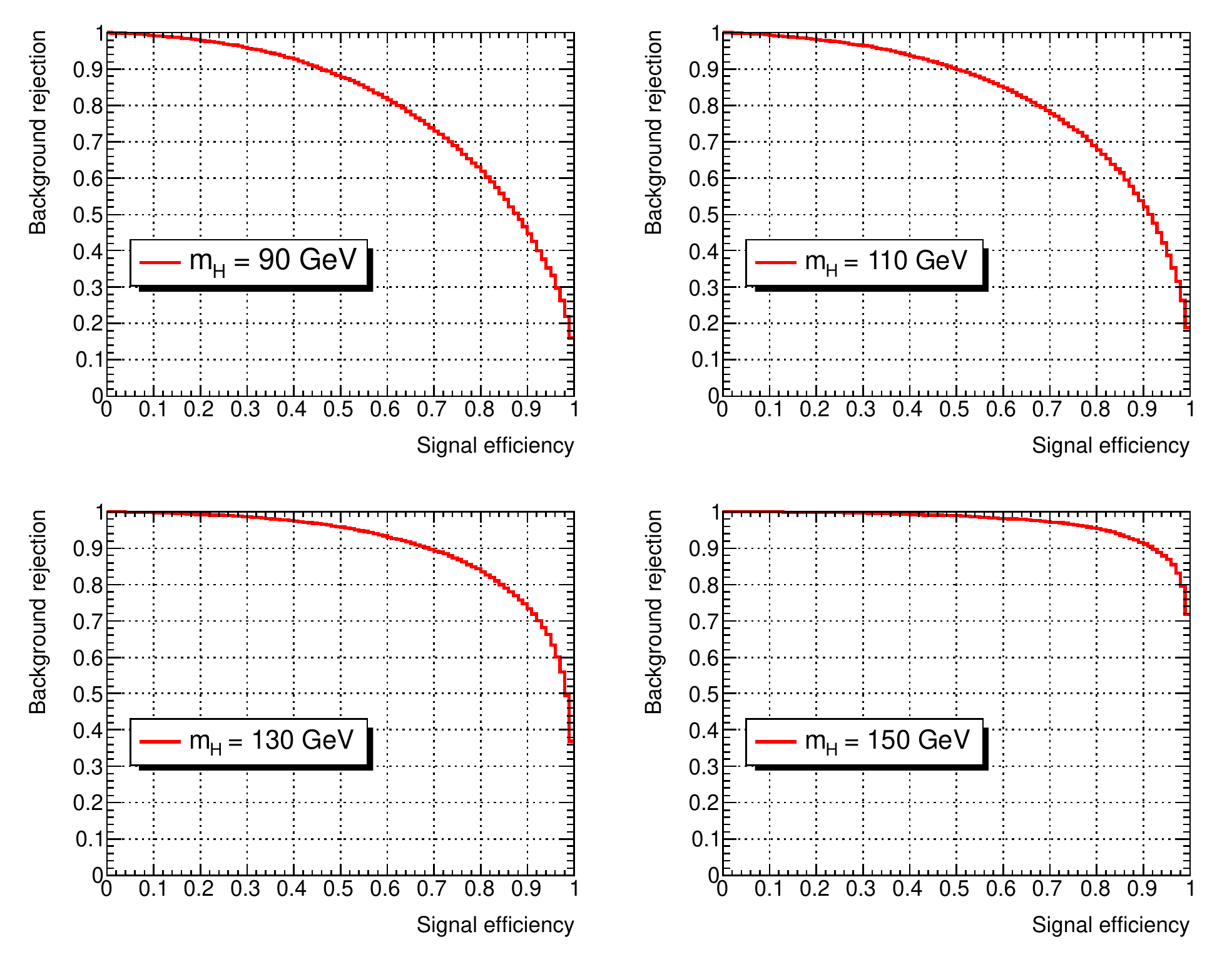}
\caption{\label{fig:roc-ttbar}
SM background rejection vs. charged Higgs signal efficiency for the four charged Higgs masses
indicated on the figure from the process $p p \to t\bar{t} X$, with $\sqrt{s}=14$ TeV.}
\end{figure}

The corresponding background rejection vs. signal efficiency curves from the process $pp \to t\bar{t}X$
calculated from the previous BDTD response at $\sqrt{s}=14$ TeV
are shown in Fig.~\ref{fig:roc-ttbar} for the four charged Higgs masses, as indicated on the frames.
For a signal efficiency value of 90\%, the background rejection varies between 50\% and 90\% as we
move from $m_{H^+}=90$ GeV to $m_{H^+}=150$ GeV.

In order to calculate the significance of our signal, we do the following simplified calculation.
We consider the less preferred case for $\tan \beta=10$, for which the branching ratio
${\cal B}(t \to H^+b)$ in the MSSM shows a dip, with ${\cal B}(t \to H^+b) \simeq 0.02$ for
 $m_{H^+}=90$ GeV (see, Fig.~\ref{fig:fig1b-higgs.pdf}).
For the process $pp \to t\bar{t}X$, the trigger is based on the decay $t \to b W^+ \to b \ell^+\nu_\ell$, with
$\ell^+=e^+,\mu^+$, which has a summed branching ratio of about 0.2.
 Since, in the large-$\tan \beta$ limit
we are working, ${\cal B}(H^+ \to \tau^+\nu_\tau) \simeq 1$, and the $\tau^+$-decay mode we are concentrating on
is $\tau^+ \to \rho^+ \bar{\nu}_\tau$, which has a branching ratio of 0.25, the product branching ratio
 $t \to b H^+ \to b (\tau^+ \nu_\tau) \to b(\rho^+ \bar{\nu}_\tau)\nu_\tau=5\times 10^{-3}$, which taking
into account the trigger is reduced to $1.0 \times 10^{-3}$. For an integrated luminosity of 10 (fb)$^{-1}$
at $\sqrt{s}=14$ TeV, and summing over the charge conjugated modes yielding a factor 2,
 this yields $2 \times10^{4}$ signal events. For the background events, resulting from the
production and the SM decays from the process $pp \to t\bar{t}X$, the corresponding product branching ratio
is 2.5\%, which together with the trigger branching gives $5\times 10^{-3}$, resulting in
$10^{5}$ background events, where we have again taken into account the factor 2 from the sum of the
charge conjugated states. Using the BDTD analysis, we get for a 50\% signal efficiency, a
background rejection of 90\%. Thus, our estimated significance will be 
\begin{equation}
S= \frac{N_{\rm signal~events}}{\sqrt{N_{\rm background~events}}}= \frac{10^{4}}{\sqrt{ 10^4}}\simeq 100~.
\end{equation}
A more realistic calculation should consider a factor of 2 reduction due to the acceptance cuts, discussed
in section A, as well as the efficiency to tag two $b$-jets which is another factor of 2, and the efficiency
of reconstructing a $\tau-{\rm jet}$, estimated as 0.3~\cite{Aad:2009wy}.
 This amounts to a factor of about 10 reduction in
both the number of signal and background events, resulting in a significance of about 30. This is high enough
to take another factor 2 reduction due to various other cuts, which will be inevitable in a detector-based
analysis taking into account non-$t\bar{t}$ backgrounds, not estimated here.
Of course, this significance goes down as $m_{H^+}$ increases, keeping $\tan \beta$ fixed. Thus, for example,
for $\tan \beta=10$ and $m_{H^+}=150$ GeV, the reduction in the number of events will be approximately 5
(a factor 10 decrease in ${\cal B}(t \to H^+b)$, compensated by a factor 2 increase in the signal efficiency
calculated from the BDTD response). This would yield $S\simeq 6$, which is just above the discovery limit 
for a charged Higgs below the top quark mass.

\begin{figure}[ht]
\centering
\hspace*{-2cm}\includegraphics[width=14.5cm,height=10.5cm]{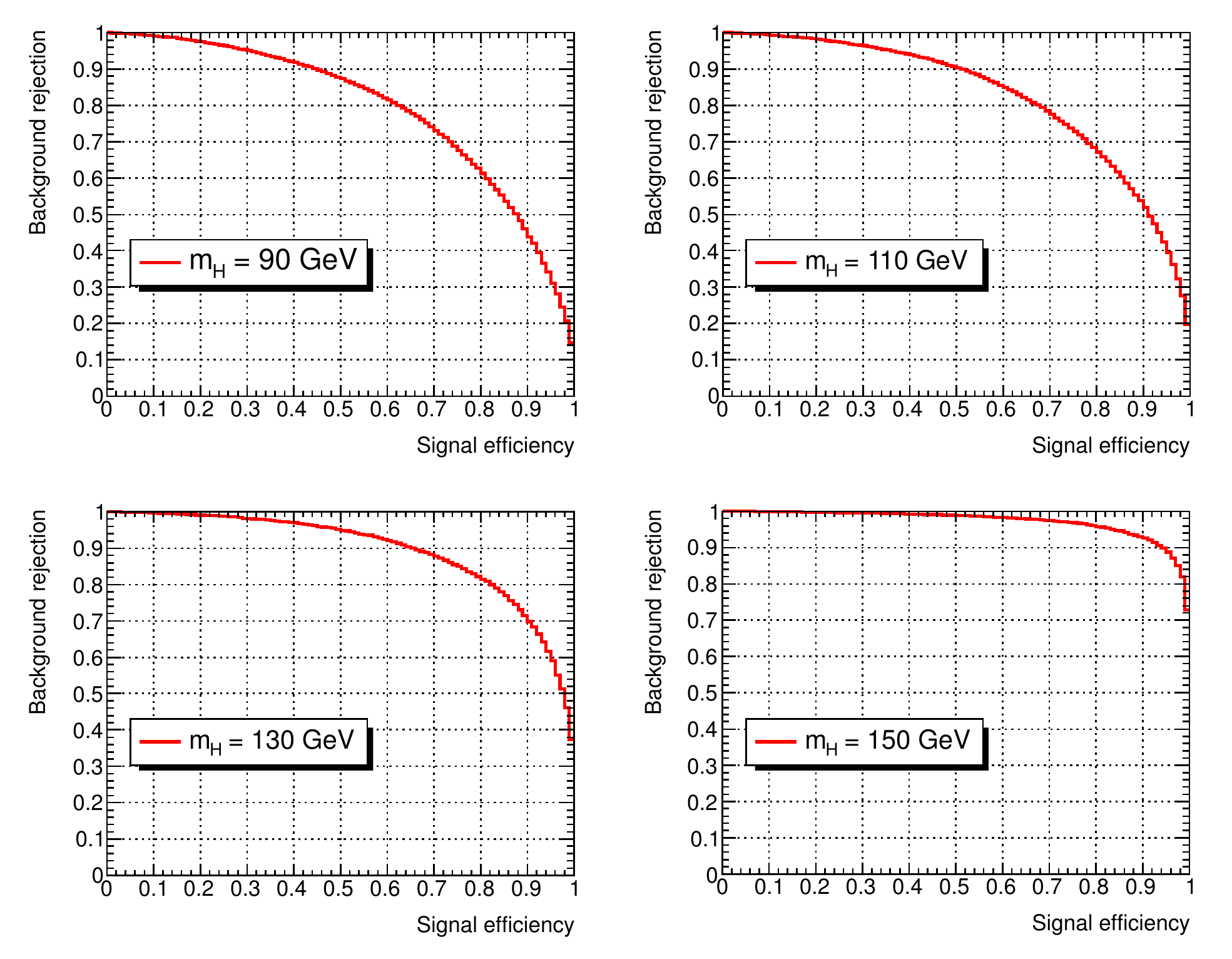}
\caption{\label{fig:roc-ttbar-pt20}
SM background rejection vs. charged Higgs signal efficiency for the four charged Higgs masses
indicated on the figure from the process $p p \to t\bar{t} X$. This figure differs from the
one shown in Fig.~\ref{fig:roc-ttbar} in the minimum transverse momentum of the $\tau$-jet,
which is set to 20 GeV as opposed to 10 GeV used in the earlier figure.}
\end{figure}

A number of checks has been performed in order to test the robustness of the results. 
For instance, the cut on the minimum transverse momentum of the $\tau$-jet has been raised from 
10 GeV to 20 GeV. The corresponding figure displaying the background rejection vs.~ the
charged Higgs signal efficiency is shown in Fig.~\ref{fig:roc-ttbar-pt20}. A comparison with
Fig.~\ref{fig:bdtd-ttbar}, obtained with a 10 GeV cut on the  minimum transverse momentum of the
$\tau$-jet, shows that the two figures are very similar. The price to pay for the acceptance is,
relatively speaking, minor, going down from 0.6 to 0.5. We had conservatively taken this to be
0.5 in our numerical calculations. 

\begin{figure}[ht]
\centering
\hspace*{-2cm}\includegraphics[width=14.5cm,height=10.5cm]{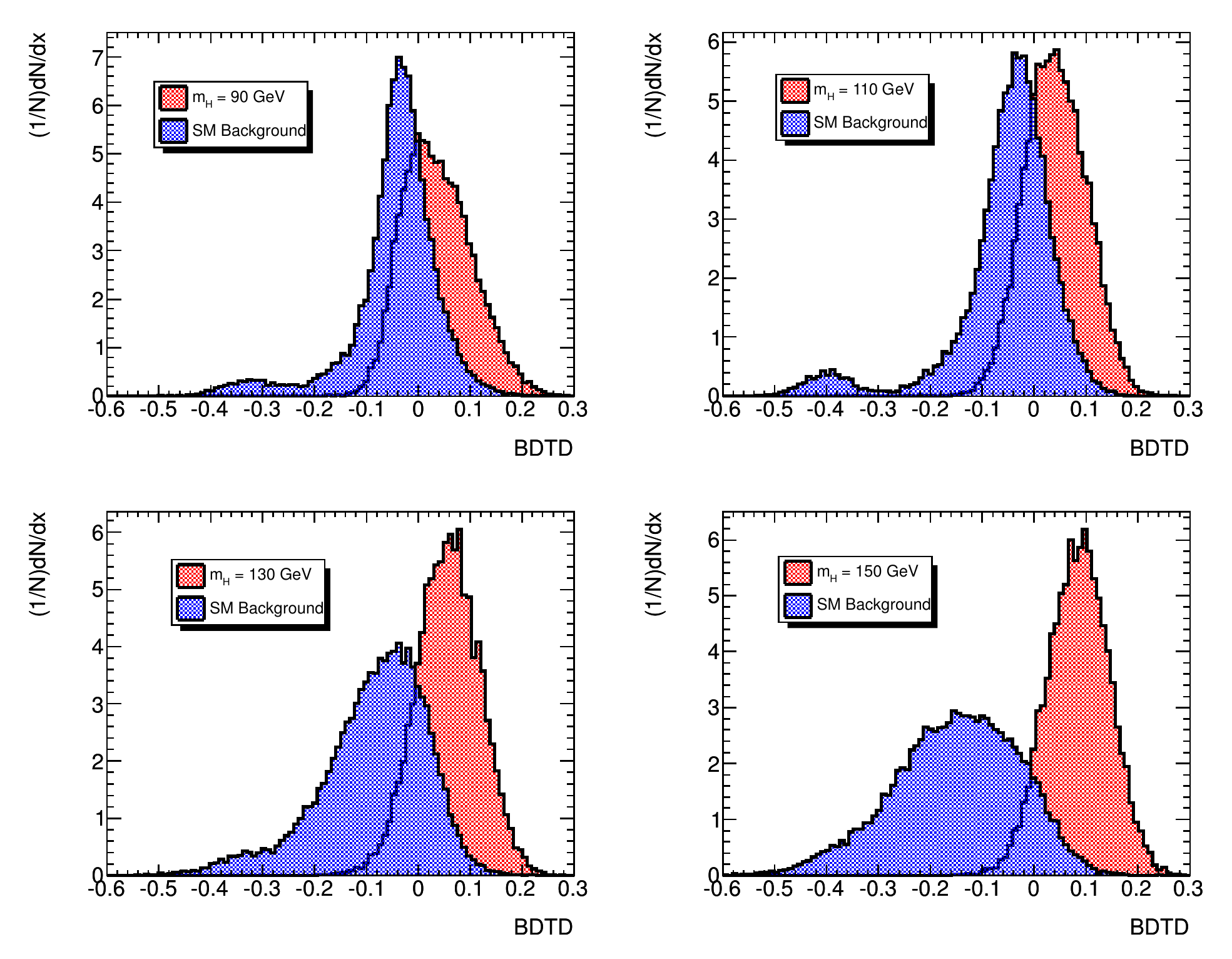}
\caption{\label{fig:bdtd-ttbar-7Tev}
BDTD response functions for $p p \to t\bar{t} X$, with $\sqrt{s}=7$ TeV, followed by the decay
 $t \to W^+ b$ (SM) and the decay chain $t \to H^+ b$, with
 the SM background (in shaded blue) and the charged Higgs signal
process (in shaded red) for four different charged Higgs masses.}
\end{figure}
\begin{figure}[ht]
\centering
\hspace*{-2cm}\includegraphics[width=14.5cm,height=10.5cm]{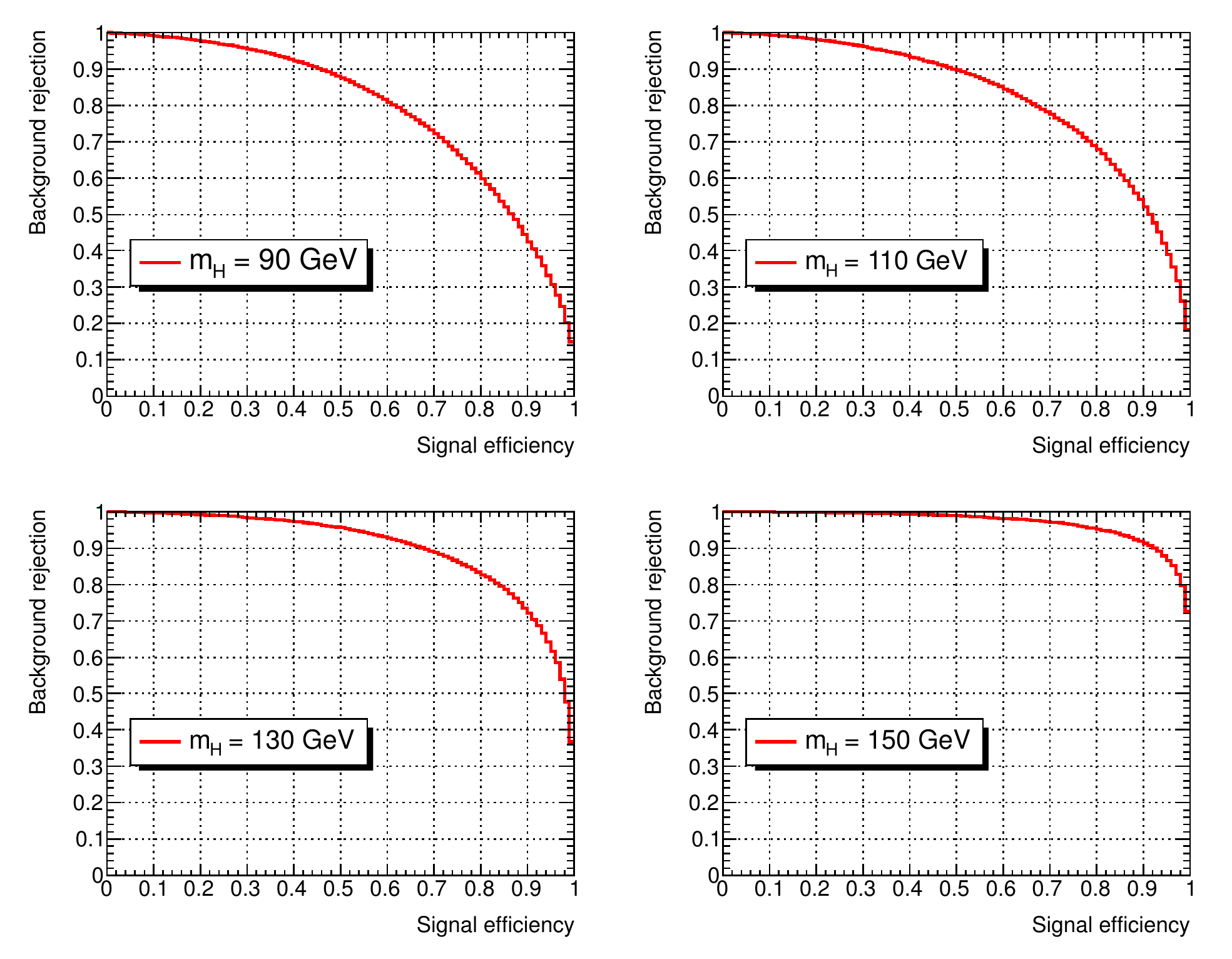}
\caption{\label{fig:roc-ttbar-7TeV}
SM background rejection vs. charged Higgs signal efficiency for the four charged Higgs masses
indicated on the figure from the process $p p \to t\bar{t} X$ with $\sqrt{s}=7$ TeV.}
\end{figure}

The above analysis presented for the LHC energy $\sqrt{s}=14$ TeV has been repeated for a
center of mass energy $\sqrt{s}= 7$ TeV, at which energy the LHC is collecting data currently.
As of preparing this report, the integrated luminosity of the LHC is above 1 inverse femtobarn,
and the projection for end 2012 is of order 10 inverse femtobarns. We have generated events at
$\sqrt{s}=7$ TeV, and have calculated all the distributions presented earlier for 14 TeV.
The shapes of these distributions are essentially similar. This is reflected in the BDTD response
functions for the SM background and the charged Higgs signal, presented in Fig.~\ref{fig:bdtd-ttbar-7Tev},
and in the SM background rejection vs. the charged Higgs signal efficiency, shown in
Fig.~\ref{fig:roc-ttbar-7TeV}. However, the cross sections for $p p \to t\bar{t} X$ at 7 TeV is
approximately a factor 4 smaller than at 14 TeV~\cite{Kidonakis:2008mu,Langenfeld:2009tc}. This
implies that our calculations for the significance obtained at $\sqrt{s}= 14$ TeV have to be
divided by a factor 2 to get the corresponding significance at $\sqrt{s}= 7$ TeV. This will reduce the
sensitivity of the charged Higgs in $\tan \beta$- $m_{H^+}$ plane. For example, for
$m_{H^+}$ close to the kinematic limit $m_t-m_b$, a signal is expected only for $\tan \beta > 20$.

A potential dilution of the polarization information has to be kept in mind. The single charged-prong
hadronic decays of the $\tau^\pm$ are essentially made up of the decays $\tau^\pm \to \pi^\pm +\nu_\tau$
(with a branching ratio of 10.9\%), $\tau^\pm \to \rho^\pm (\to \pi^\pm \pi^0) +\nu_\tau$ decays
(with a branching ratio of 25.5\%) and $\tau^\pm \to a_1^\pm (\to \pi^\pm \pi^0 \pi^0) +\nu_\tau$ decays
(with a branching ratio of 9.3\%). Separating the $\pi^\pm \nu_\tau$ mode from the $\rho^\pm \nu_\tau$
mode should, in principle, be possible due to the lack of deposited energy 
in the $\pi^0$ or electromagnetic cluster accompanying the $\pi^\pm$
in the former, but separating the $\rho^\pm \nu_\tau$ mode from the $a_1^\pm \nu_\tau$ mode will not
be easy. Fortunately, the branching ratio of the latter is only 40\% of the former. So, the number of 
$\pi^0$ clusters (0, 1 and 2) will have to be included in the analysis as a new variable. The
$\tau^\pm \to \rho^\pm \nu_\tau$ decays  dominate the one charged track ($\pi^\pm, K^\pm$) and
one electromagnetic or $\pi^0$ cluster. However, we stress that the BDT can be trained to reduce the
dilution.

In a realistic analysis, a further source of reduction in our estimates of the significance would come from
the wrong assignment of the $b$-jet charges, though this effect is minor compared to the ones discussed above.
The $b$-jet charge identification efficiency is estimated at present to be around 65\%~\cite{Aad:2009wy},
using standard techniques based on a weighted average of the charges of the particles in the jet, with
the weights being proportional to their momenta. However, a simple algorithm can be designed, which takes
into account in addition the angular correlations between the trigger lepton, the tau-jet and the charges,
reducing the $b$-jet mis-assignment to about 20\% for the charged Higgs masses close to the $W^\pm$ mass.
For higher charged Higgs masses, this can be further brought down by simply taking the $b$-jet with the
smaller (larger) transverse momentum to be that associated with the charged Higgs (resp.~$W^\pm$) boson.

We also mention that we have not considered the background from the process
 $pp \to t\bar{t} \to (b \ell \nu_\ell) (b jj)$. However, it has been shown in~\cite{Aad:2009wy}
that this background can be well separated in a standard cut analysis from the
$pp \to t \bar{t} \to (b \ell \nu_\ell) (b \tau \nu_\tau)$ process. With our TMVA approach, this
background will be tamed though we will have to introduce also the  missing $E_T$ as a variable
in the BDT training.
 We are aware of the non-$t\bar{t}$ background, which are dominated by the $Z +{\rm jets}$ and
  $W +{\rm jets}$. These have been studied in great detail in 
\cite{ATLAS-2010-006}, with the conclusion that they can be brought below the signal by the additional use
of the $E_T^{\rm miss}$-cut. We have not used the $E_T^{\rm miss}$-cut, as we have concentrated only
on the SM $t\bar{t}X$ background, but will do so in a more realistic detector-based analysis in the
future.

\section{Single $t/\bar{t}$ production and the decay chains
 $t \to b W^+/H^+ \to b (\tau^+ \nu_\tau)$ at the LHC}

\subsection{Cross sections at the LHC}
The single top (or anti-top) cross sections in hadron hadron collisions have been
calculated in the NLO
 approximation~\cite{Harris:2002md,Cao:2004ap,Heim:2009ku,Kidonakis:2006bu,Kidonakis:2007ej}.
 Recalling that there are 
three basic processes at the leading order which contribute to
 $\sigma (p\bar{p} \to t/\bar{t} X)$, namely the $t$-channel: $qb \to q^\prime t$,
the $s$-channel: $q\bar{q}^\prime \to \bar{b}t$; and the associated $tW$ production
$bg \to tW^-$, the cross section estimated at the Tevatron is~\cite{Kidonakis:2009sv}:
  $\sigma (p\bar{p} \to t X)=\sigma(p \bar{p} \to \bar{t}X) \simeq 1.8$ pb for both the
top and anti-top production.
 At the LHC@14 TeV, one estimates  $\sigma(pp \to tX) \simeq 200$ pb and about half this number for
$\sigma(pp \to \bar{t}X)$, yielding the summed single top and anti-top cross sections at
about 300 pb, also approximately two orders of magnitude larger
than those at the Tevatron. With a luminosity of 10 fb$^{-1}$, one anticipates 
$O(3 \times 10^6)$ single top (or anti-top) events.

\begin{figure}[ht]
\includegraphics[width=9.5cm,height=6.5cm]{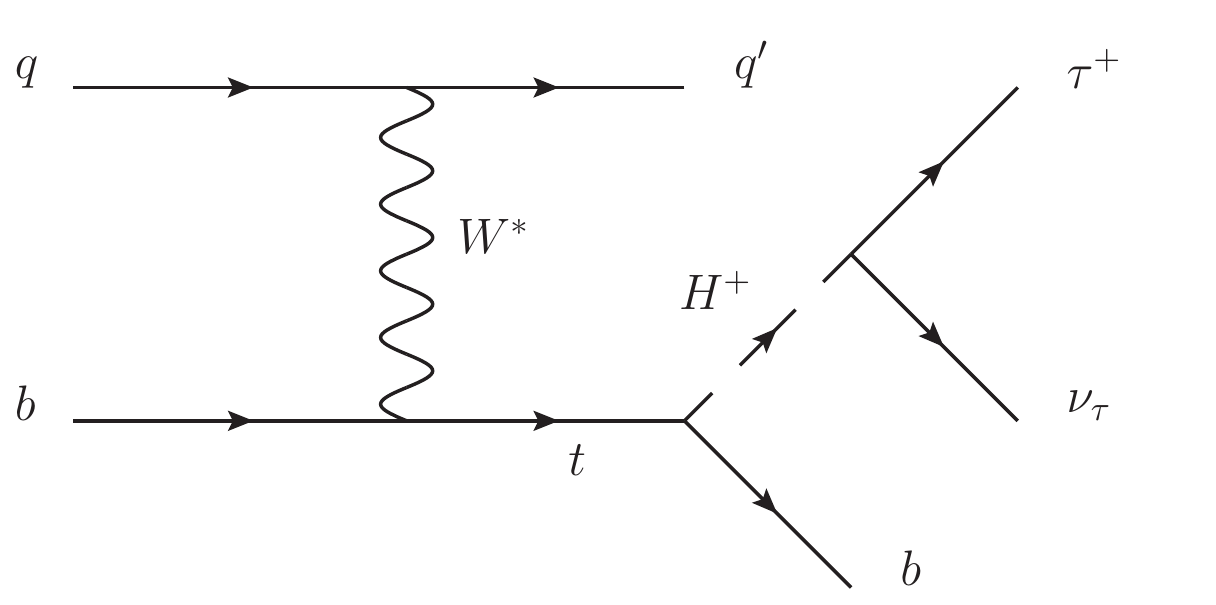}
\caption{\label{fig:feyn-singlet}
Feynman diagram for $q b \to q^\prime t$, followed by the decay $t \to b (H^+ \to \tau^+ \nu_\tau)$.}
\end{figure}

As mentioned in the introduction, there are three different mechanisms of producing a single top
(or anti-top) quark in hadronic collisions, the $s$-channel, the $t$-channel, and the associated
production $tW$-channel. The Feynman diagram for the dominant $t$-channel partonic process 
 $q b \to q^\prime t$, followed by the decay $t \to b (H^+ \to \tau^+ \nu_\tau)$ is shown in
Fig.~\ref{fig:feyn-singlet}. The partonic cross section is then convoluted with the parton distribution
functions to calculate the cross sections in $pp \to t +X$ and $pp \to \bar{t} + X$. Since, we are
using PYTHIA 6.4~\cite{Sjostrand:2006za} to do the simulation of the single top (or anti-top) production,
not all channels are encoded there yet. However, as we use the generator to calculate the acceptance only,
but the total cross sections are normalized to the theoretical calculations, the estimates presented
here should hold approximately. Since most of the distributions calculated by us for
the processes $ p p \to t\bar{t} x$ and $ pp \to t/\bar{t} X$ are in the same variables, we
comment only briefly on the distributions for the signal $t \to bH^+ \to b \tau^+\nu_\tau$ and  
the background process $t \to bW^+ \to b \tau^+\nu_\tau$.

\begin{figure}[ht]
\centering
\includegraphics[width=0.99\textwidth,height=0.5\textwidth]{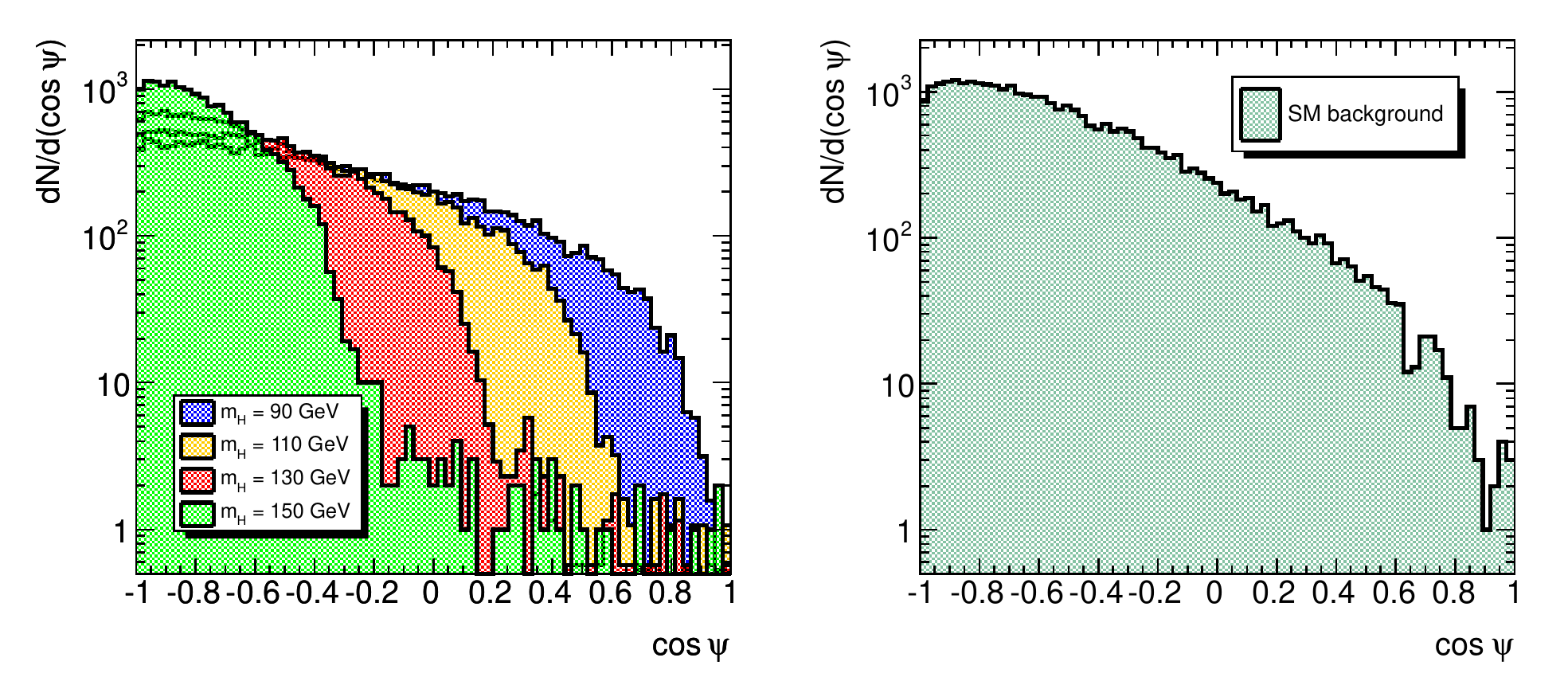}
\caption{\label{fig:cos-psi-singlet}
The distribution $dN/d \cos \psi$ for the $pp \to t/\bar{t} +X$ production as measured in 
the decay chain
$t \to b W \to b (\tau \nu_\tau) \to b (\rho \bar{\nu}_\tau) \nu_\tau)$ (right-hand frame),
and in $t \to b H \to b (\tau \nu_\tau) \to b (\rho \bar{\nu}_\tau) \nu_\tau)$ for four
different charged Higgs masses, as indicated on the figure (left-hand frame).}
\end{figure}

In Fig.~\ref{fig:cos-psi-singlet}, we show the
distribution $dN/d \cos \psi$ for the $pp \to t/\bar{t} +X$ production as measured in 
the decay chain for the SM background process
$t \to b W \to b (\tau \nu_\tau) \to b (\rho \bar{\nu}_\tau) \nu_\tau)$ (right-hand frame),
and for the signal  $t \to b H \to b (\tau \nu_\tau) \to b (\rho \bar{\nu}_\tau) \nu_\tau)$
(left-hand frame) for four different charged Higgs masses, as indicated on the figure.
The SM background in the process $pp \to t/\bar{t} +X$ falls more steeply as a function of
$\cos \psi$ than is the case for the $t\bar{t}$ production $pp \to t \bar{t} +X$, due to the
acceptance cuts. The trend is similar in the signal process. However, also in the single top
(or anti-top) production, this distribution provides a good discriminant as input to the BDTD analysis.

\begin{figure}[ht]
\centering
\hspace*{-2cm}\includegraphics[width=14.5cm,height=10.5cm]{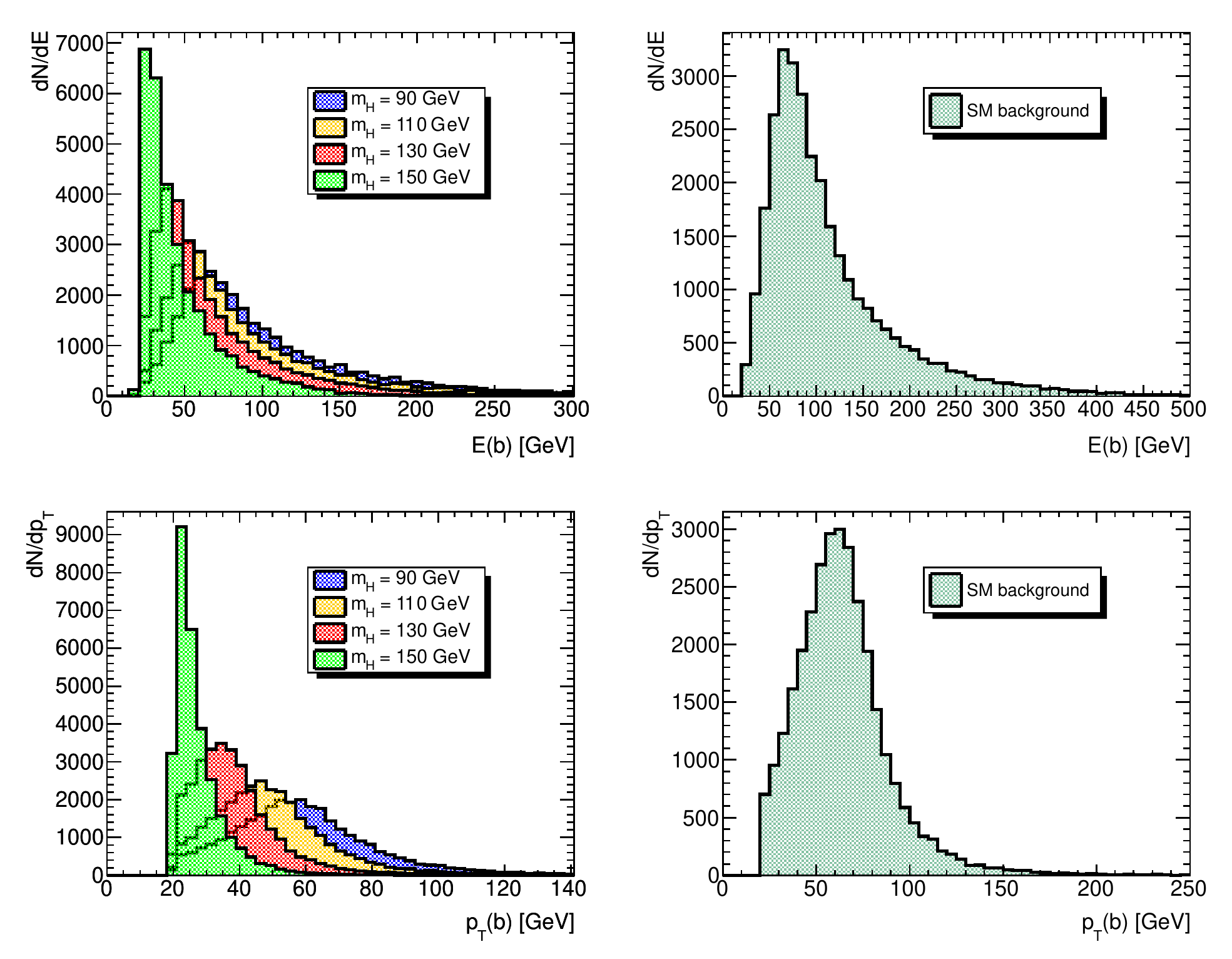}
\caption{\label{fig:bjets-singlet}
Distributions in the energy of the $b$-jet, $E(b)$, and transverse momentum of the $b$-jet, $p_T(b)$
 from the
process $p p \to t/\bar{t} X$, followed by the decay $t \to W^+ b$ (right-hand frames), and
the same distributions for the decay chain $t \to H^+ b$ with the four indicated charged Higgs masses
(left-hand frame).}
\end{figure}

The distributions  in the energy of the $b$-jet, $E(b)$, and transverse momentum of the $b$-jet, $p_T(b)$
 from the 
process $p p \to t/\bar{t} X$, followed by the SM decay $t \to W^+ b$
are shown in Fig.~\ref{fig:bjets-singlet} (right-hand frames), and
the same distributions for the decay chain $t \to H^+ b$ with the four indicated charged Higgs masses
are also shown in this figure (left-hand frame). These distribution are very similar to the ones shown for the
$pp \to t\bar{t} X$ processes, as they essentially reflect the kinematics of the decays $t \to W^+ b$ and $t \to H^+ b$.

\begin{figure}[ht]
\centering
\hspace*{-2cm}\includegraphics[width=14.5cm,height=10.5cm]{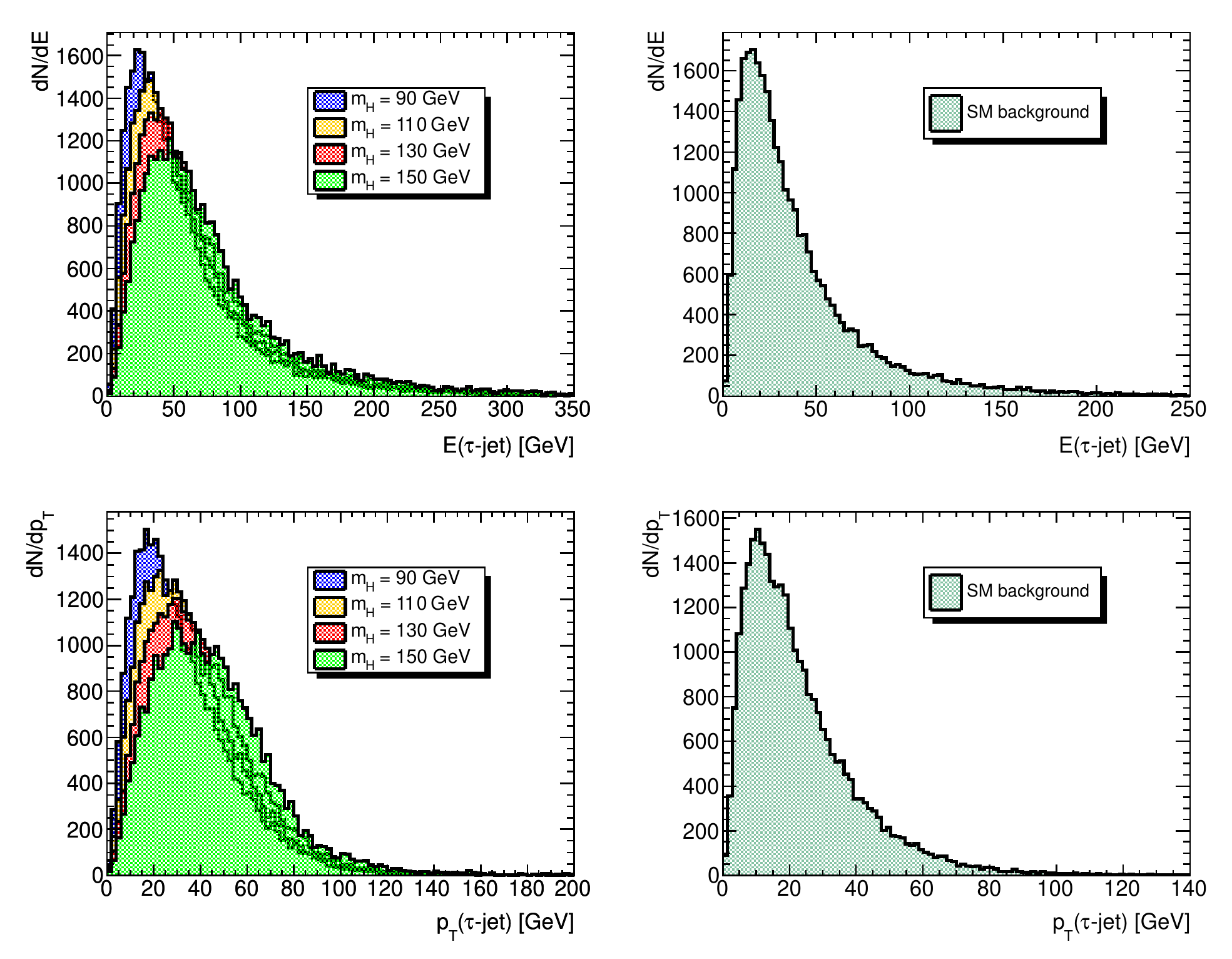}
\caption{\label{fig:tau-singlet}
Distributions in the energy of the $\tau$-jet, $E(\tau-{\rm jet})$, and transverse momentum of the
 $\tau$-jet, $p_T(\tau-{\rm jet}) $ from the
process $p p \to t/\bar{t} X$, followed by the SM decay $t \to W^+ b$ (right-hand frames), and
the same distributions for the decay chain $t \to H^+ b$ with the four indicated charged Higgs masses
(left-hand frame).}
\end{figure}

In Fig.~\ref{fig:tau-singlet}, we show the corresponding distributions for the $\tau$-jet, $E(\tau-{\rm jet})$,
 and for the transverse momentum of the
 $\tau$-jet, $p_T(\tau-{\rm jet}) $ from the
process $p p \to t/\bar{t} X$, followed by the SM decay $t \to W^+ b$ (right-hand frames), and
the same distributions for the decay chain $t \to H^+ b$ with the four indicated charged Higgs masses
(left-hand frame). These distributions, likewise, are very similar to the ones shown for the $t\bar{t}$
production case, shown in the previous section.

\begin{figure}[ht]
\centering
\hspace*{-2cm}\includegraphics[width=14.5cm,height=10.5cm]{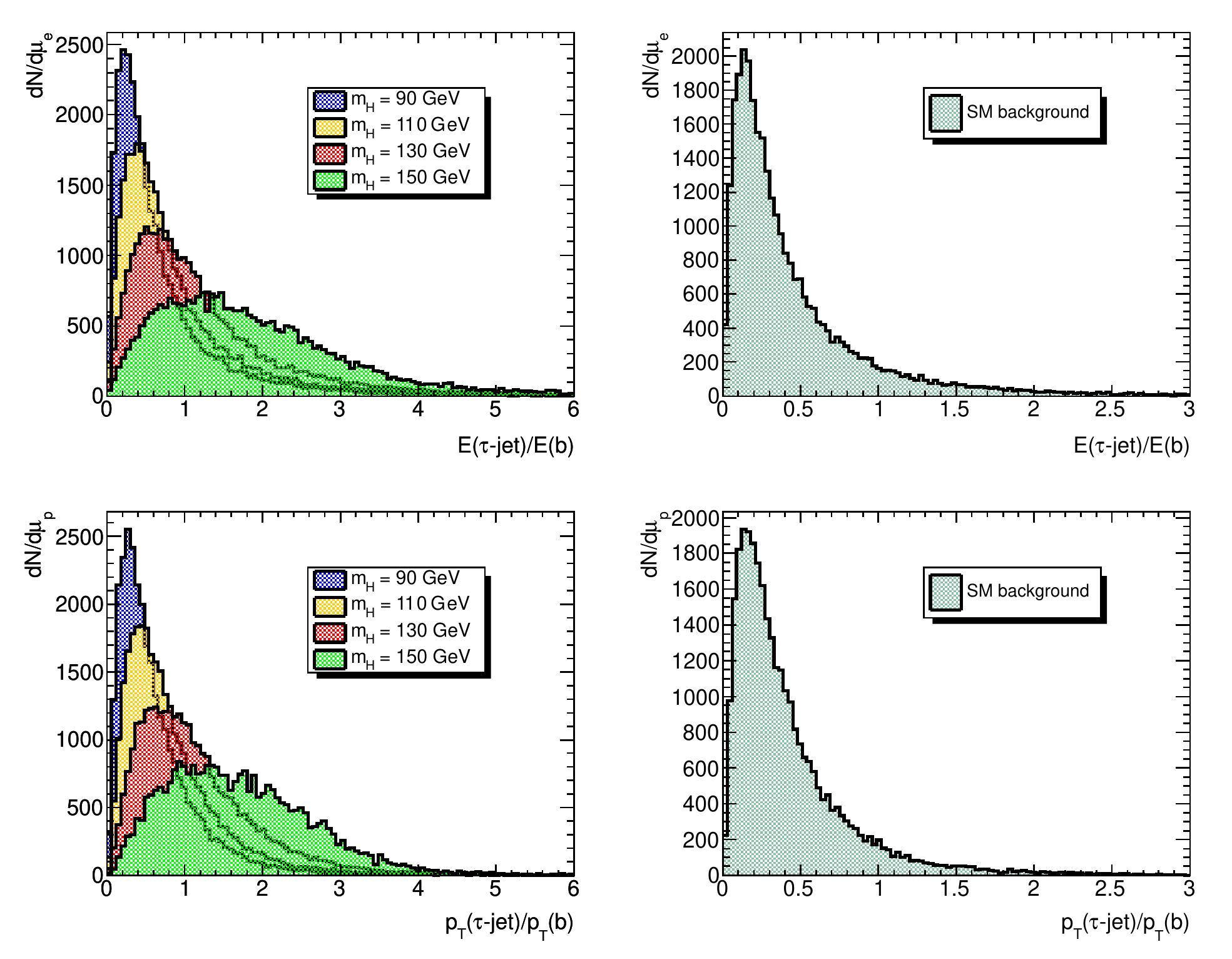}
\caption{\label{fig:tauB-singlet}
Distributions in the ratio $E(\tau-{\rm jet})/E(b)$ and $p_T(\tau-{\rm jet})/p_T(b)$
from the process $p p \to t/\bar{t} X$, followed by the SM decay $t \to W^+ b$ (right-hand frames), and
the same distributions for the decay chain $t \to H^+ b$ with the four indicated charged Higgs masses
(left-hand frame).}
\end{figure}

The distributions in the ratio $E(\tau-{\rm jet})/E(b)$ and $p_T(\tau-{\rm jet})/p_T(b)$
from the process $p p \to t/\bar{t} X$, followed by the SM decay $t \to W^+ b$
are shown in Fig.~\ref{fig:tauB-singlet} (right-hand frames), and
the same distributions for the decay chain $t \to H^+ b$ are also shown in this
figure  with the four indicated charged Higgs masses
(left-hand frame). As anticipated, these distributions are also similar in the single top (anti-top)
production and in the $t\bar{t}$ production.

\begin{figure}[ht]
\centering
\hspace*{-2cm}\includegraphics[width=14.5cm,height=10.5cm]{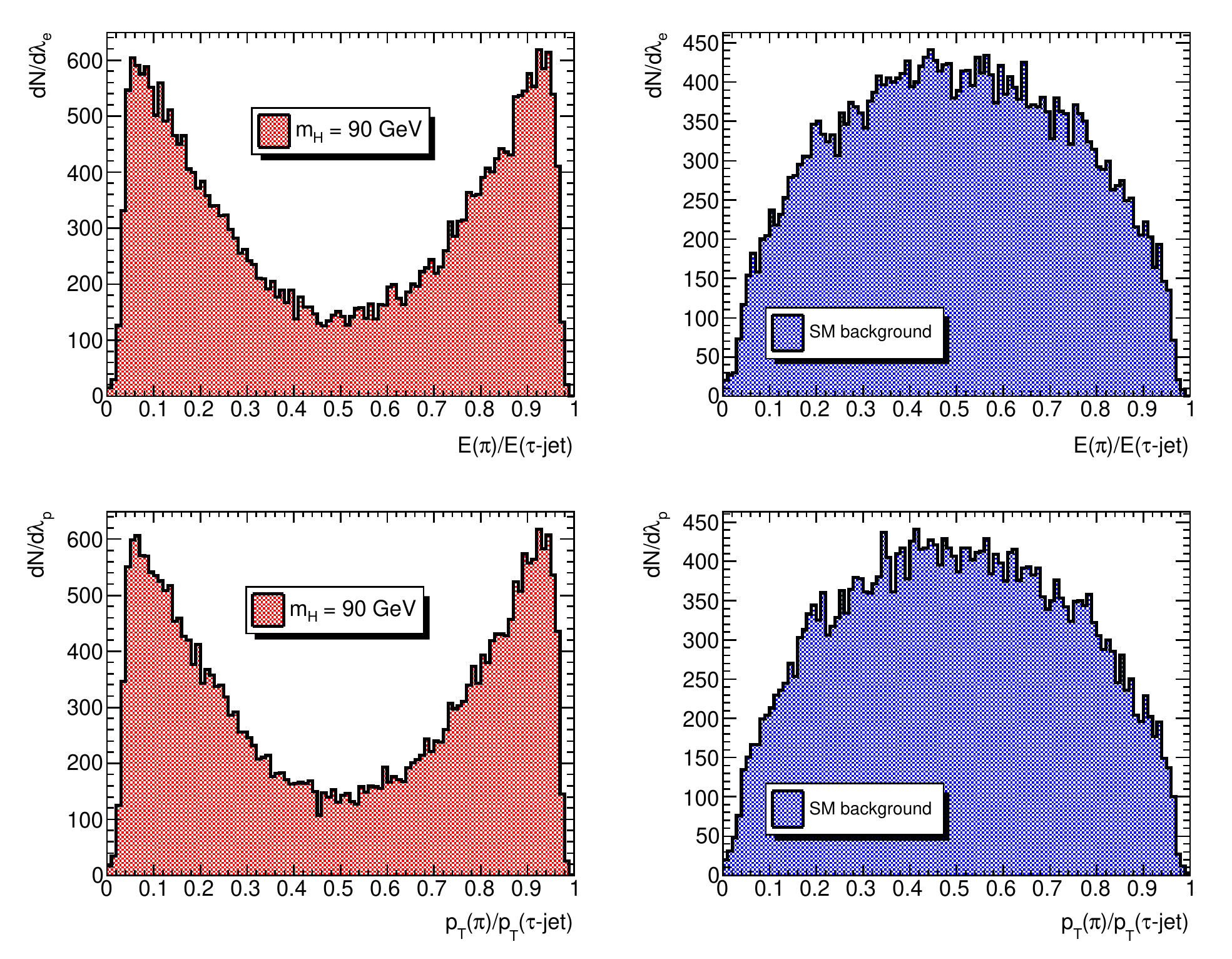}
\caption{\label{fig:polarization-singlet}
Distributions in the fractional energy of the single-charged prong ($\pi^+$ in $\tau^+$-jet),
$E(\pi)/E(\tau-{\rm jet})$, and in the transverse
momentum of the single-charged prong, $p_T(\pi)/p_T(\tau-{\rm jet})$ from the
$p p \to t/\bar{t} X$, followed by the SM decay $t \to W^+ b$ (right-hand frames), 
and the same distributions for the decay chain $t \to H^+ b$ with the four indicated charged Higgs masses
(left-hand frame).}
\end{figure}

The effects of different chiralities in the SM decay chain $t \to b W^+ \to b (\tau^+ \nu_\tau)$ followed by
the $\tau^+$ decay $\tau^+ \to \rho^+ \bar{\nu}_\tau$, and in the signal process 
$t \to b H^+ \to b (\tau^+ \nu_\tau)$ followed by
the $\tau^+$ decay $\tau^+ \to \rho^+ \bar{\nu}_\tau$ are shown in Fig.~\ref{fig:polarization-singlet}.
Once again, these distributions in the fractional energy of the single-charged prong ($\pi^+$ in $\tau^+$-jet),
$E(\pi)/E(\tau-{\rm jet})$, and in the transverse
momentum of the single-charged prong, $p_T(\pi)/p_T(\tau-{\rm jet})$
are very similar in the processes
$p p \to t/\bar{t} X$ and $pp \to t\bar{t} X$, as expected. 

\begin{figure}[ht]
\centering
\hspace*{-2cm}\includegraphics[width=14.5cm,height=10.5cm]{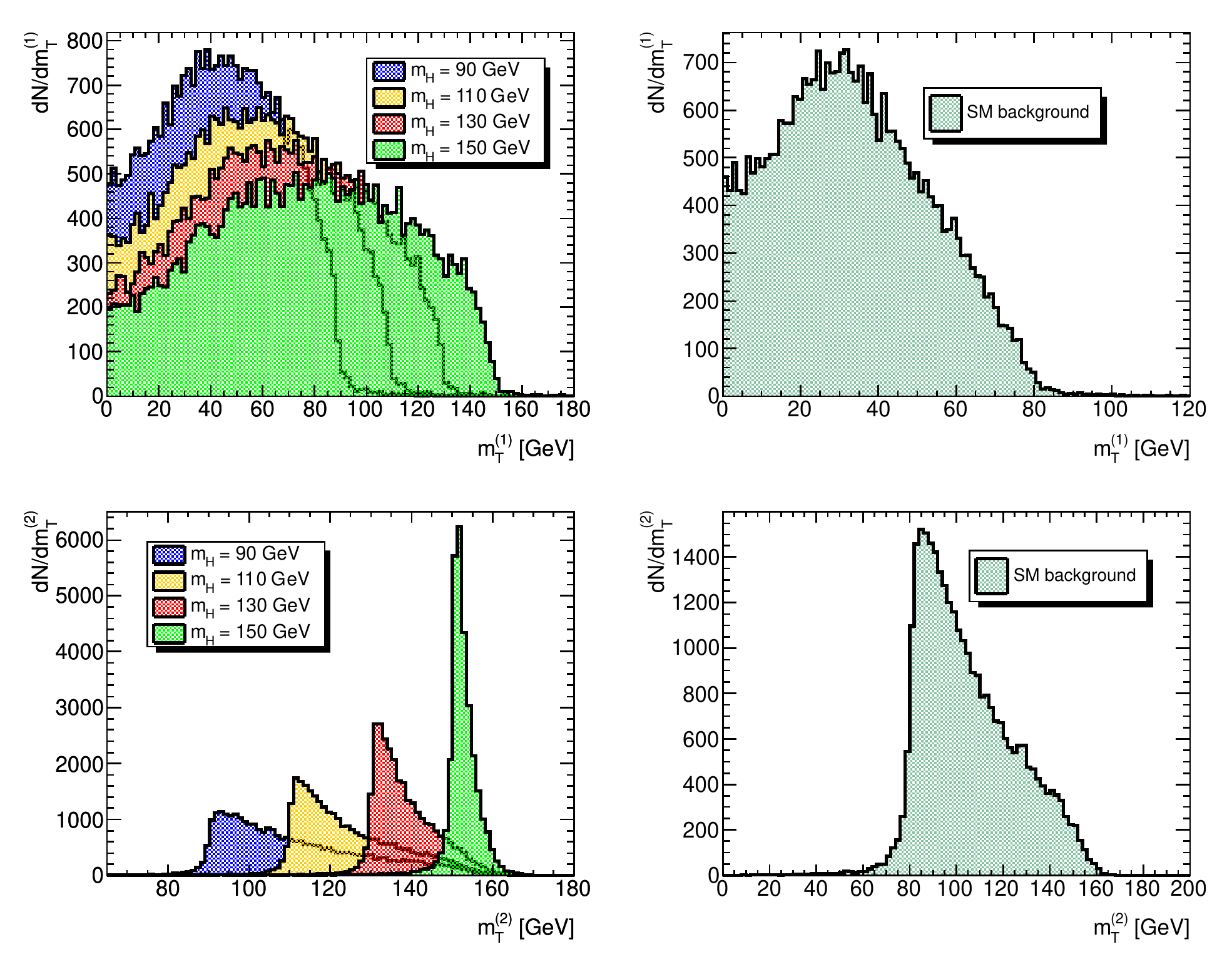}
\caption{\label{fig:mass-singlet}
Transverse mass distributions for the $W^\pm$ in the process
$pp \to t/\bar{t} +X$ followed by the decay $t \to b W^+$
 (right-hand frames) and for the $H^\pm$ transverse mass for the decay chain
$t \to H^+ b$ with the four indicated charged Higgs masses
(left-hand frame). The definitions used for defining the transverse masses $m_T^{(1)}$
and $m_T^{(2)}$ are given in the text.}
\end{figure}

One important difference between the analysis of the single top (or anti-top) production compared to the
$t\bar{t}$ production process lies in the fact that the missing transverse energy and momentum can be
ascribed in the former to the $\tau$-neutrino, $\nu_\tau$. This is different in the case of the
$t\bar{t}$ production, as  one of the $t$-or $\bar{t}$-quarks decays via $t \to b W^+ \to b \ell^+ \nu_\ell$, which is
used as a trigger. Thus, the missing transverse energy or momentum can not be traced to the decay of the
$\tau$-lepton alone in the case of $t\bar{t}$ production. As already stated in~\cite{Gross:2009wia}
the missing transverse energy and momentum profile
in the case of the single top (or anti-top) process $pp \to t/\bar{t}X$ followed by $t \to b  H^+  \to b \tau^+ \nu_\tau$
can be used to constrain the mass of the charged Higgs. We pursue this idea, by  using two different definitions of
the transverse mass. In the first case, called $m_T^{(1)}$, this is defined as
in~\cite{Smith:1983aa}:
\begin{equation}
m_T^2 = 2 p_T^\ell p_T^\nu (1 -\cos \phi_{\ell \nu})~,
\end{equation}
where $p_T^\ell$, $ p_T^\nu$, and $\phi_{\ell \nu}$ are the momenta and angle between the leptons in the plane
perpendicular to the $pp$ collision axis. This definition was proposed to determine the transverse mass of the
$W^\pm$ boson in $p\bar{p}$ collisions using the decay modes $W^\pm \to e^\pm \nu_e$ and $W^\pm \to \mu^\pm \nu_\mu$.
In our case, where the charged Higgs decays via $H^+ \to \tau^+ \nu_\tau$,
the charged lepton is the $\tau^+$, which is not measured experimentally.  Since, we use the decay
 $\tau^+ \to \rho^+ \bar{\nu}_\tau$, we replace the $p_T^\ell$ by the $p_T$ of the $\rho^+$. The resulting 
$m_T^{(1)}$-distributions are shown in the upper two frames in Fig.~\ref{fig:mass-singlet} for the
SM background (right-hand frame) and the charged Higgs case (left-hand frame). As seen from the distributions
shown in the left-hand frame, this definition is not useful to see the Jacobian peak in the transverse mass of
the $H^\pm$. This is anticipated since there are are two undetected neutrinos from the $H^\pm$ vertex.
 The distributions in $m_T^{(1)}$ for the SM ($W^\pm$)-background and the $H^\pm$-signal
are different, and they do add to the discriminating power in the BDTD analysis.

For the processes $pp(\bar{p}) \to t\bar{t}X$ and $pp(\bar{p}) \to t\bar{b}X$, with the subsequent decay of
$t \to b(W^+,H^+)$, if one of the two $b$ jets could be associated with the semileptonic decay of the 
top quark, then the on-shell constraint for the top quark could be used in the form
 $(p^{\rm miss} + p_\ell +p_b)^2=m_{\rm t}^2$. In this case, a transverse Higgs mass can be defined
by maximizing the invariant mass, $(m_T^H)^2= {\rm max} [(p_\ell + p^{\rm miss})^2]$, since it is bounded
from above by the top quark mass, with the charged Higgs transverse mass satisfying
 $m_{H^+} \leq m_T^H \leq m_{\rm t}$, where $m_{H^+}$ is the true chraged Higgs mass. This leads to the
following transverse mass definition for $m_T^H$~\cite{Gross:2009wia}, which we call $m_T^{(2)}$,
\begin{equation}
(m_T^H)^2= \left( \sqrt{m_t^2 + (\vec{p}_T^\ell + \vec{p}_T^b + \vec{p}_T^{\rm miss})^2 } -p_T^b \right)^2
- (\vec{p}_T^\ell + \vec{p}_T^{\rm miss})^2~.
\end{equation}  
This expression holds by  neglecting the $b$-quark mass.
We have calculated the $m_T^{(2)}$ distributions, by replacing the $ \vec{p}_T^\ell$ (which is $ \vec{p}_T^\tau$
for our case) by $\vec{p}_T^\rho$. These distributions are shown in the lower two frames of
 Fig.~\ref{fig:mass-singlet}, with the SM background (yielding the Jacobian peak of the $W^\pm$) shown on
the right-hand frame, and the corresponding Jacobian peaks for the charged Higgs case, shown in the
left-hand frame. For all the four charged Higgs masses shown in this frame, the Jacobian in $m_T^{(2)}$
has a sharp peak. Measuring these distributions provides, in principle, an estimate of $H^\pm$. We
will use these distributions in $m_T^{(2)}$ to train our BDTD sample.

\begin{figure}[ht]
\centering
\hspace*{-2cm}\includegraphics[width=14.5cm,height=10.5cm]{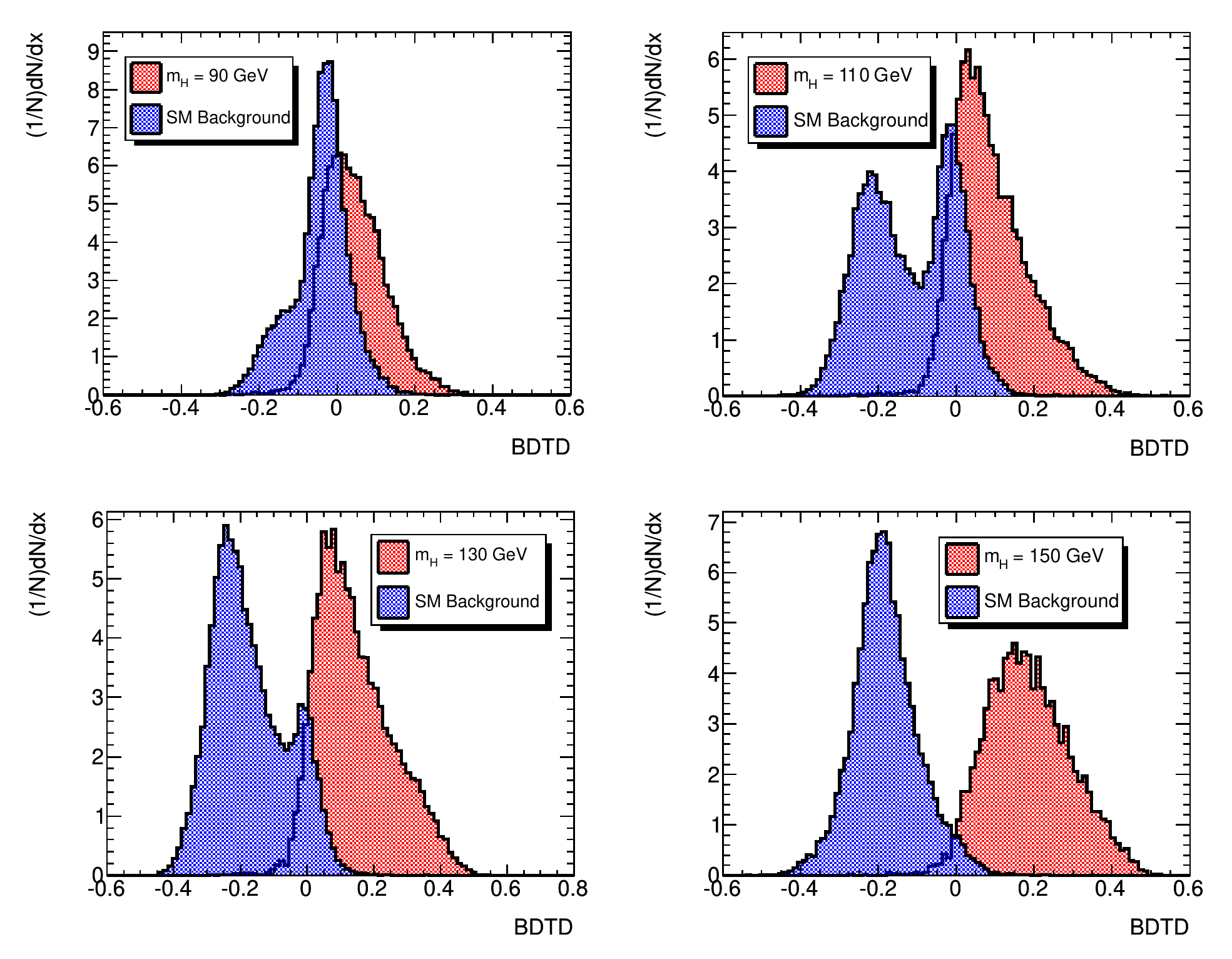}
\caption{\label{fig:bdtd-singlet}
BDTD response functions for $p p \to t/\bar{t} X$, followed by the decay
 $t \to W^+ b$ (SM) and the decay chain $t \to H^+ b$, with
 the SM background (in shaded blue) and the charged Higgs signal
process (in shaded red) for four different charged Higgs masses.}
\end{figure}

The distributions generated and discussed have been used to train the BDTD algorithms and the
resulting response functions are shown in Fig.~\ref{fig:bdtd-singlet}. The separation between the
signal and the background improves as $m_{H^+}$ increases, a trend which was also observed in the
$pp \to t\bar{t} X$ production process.

\begin{figure}[ht]
\centering
\hspace*{-2cm}\includegraphics[width=14.5cm,height=10.5cm]{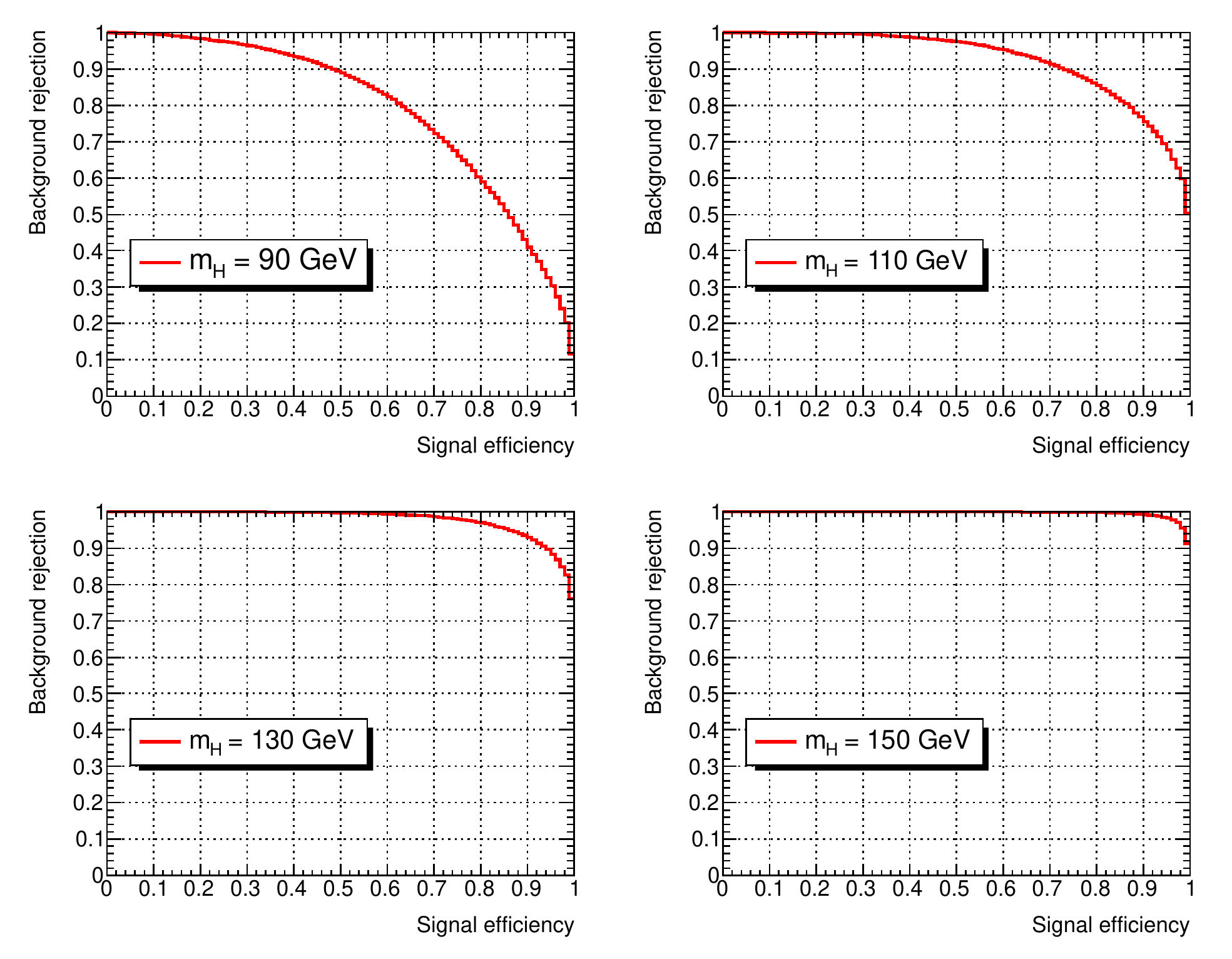}
\caption{\label{fig:roc-singlet}
SM background rejection vs. charged Higgs signal efficiency for the four charged Higgs masses
indicated on the figure from the process $p p \to t/\bar{t} X$.}
\end{figure}

The corresponding background rejection vs. signal efficiency curves from the processes
$pp \to t/\bar{t} X$ calculated from the previous BDTD response at $\sqrt{s}=14$ TeV are shown in
Fig.~\ref{fig:roc-singlet} for the four charged Higgs masses, as indicated on the frames.
For a signal efficiency value of 90\%, the background rejection varies between 40\% and 99\% as we
move from $m_{H^+}=90$ GeV to $m_{H^+}=150$ GeV.

In order to calculate the significance of our signal, we do the following simplified calculation.
We consider again the less preferred case for $\tan \beta=10$,  with ${\cal B}(t \to H^+b) \simeq 0.02$
 for $m_{H^+}=90$ GeV.
For the process $pp \to t/\bar{t}X$, our trigger is based on the $\tau$-jet, coming from the decays
$W^\pm/H^\pm \to \tau^\pm \nu_\tau$.  
 Since, in the large-$\tan \beta$ limit
we are working, ${\cal B}(H^+ \to \tau^+\nu_\tau) \simeq 1$, and the $\tau^+$-decay mode we are concentrating on
is $\tau^+ \to \rho^+ \bar{\nu}_\tau$, which has a branching ratio of 0,25, the product branching ratio
 $t \to b H^+ \to b (\tau^+ \nu_\tau) \to b(\rho^+ \bar{\nu}_\tau)\nu_\tau=5\times 10^{-3}$, which 
is the same as in the case of top-quark pair production process $pp \to t \bar{t}X$.
 For an integrated luminosity of 10 (fb)$^{-1}$, and inclusive single top cross section
$\sigma(pp \to t/\bar{t} X=300$ pb at $\sqrt{s}=14$ TeV,
 this yields $1.5 \times 10^{4}$ signal events. For the background events, resulting from the
production and the SM decays from the process $pp \to t/\bar{t}X$, the corresponding product branching ratio
is 2.5\%, again the same as in the case of top-quark pair production process $pp \to t \bar{t}X$.
yielding
$7. 5\times 10^{4}$ background events. Using the BDTD analysis, we get for a 50\% signal efficiency, a
background rejection of 90\%. Thus, our estimated significance will be 
\begin{equation}
S= \frac{N_{\rm signal~events}}{\sqrt{N_{\rm background~events}}}= \frac{7.5\times 10^{3}}{\sqrt{7.5\times 10^3}}\simeq 85~.
\end{equation}
A more realistic calculation should consider a factor of 2 reduction due to the acceptance cuts, discussed
in section A, as well as the efficiency to tag the $b$-jet, estimated as 70\%, and the efficiency
of reconstructing a $\tau-{\rm jet}$, estimated as 0.3. This amounts to a factor of about 10 reduction in
both the number of signal and background events, resulting in a significance of about 25.
Of course, this significance goes down as $m_{H^+}$ increases, keeping $\tan \beta$ fixed. Thus, for example,
for $\tan \beta=10$ and $m_{H^+}=150$ GeV, the reduction in the number of events will be approximately 5
(a factor 10 decrease in ${\cal B}(t \to H^+b)$, compensated by a factor 2 increase in the signal efficiency
calculated from the BDTD response). Since the background rejection goes up to 99\%, this would
 yield $S\simeq 25$, allowing to search for a charged Higgs in
the decay $t \to bH^+$, essentially up to a charged Higgs mass close to the kinematic limit.

We would like to stress that our philosophy in this paper is to show how to disentangle  the
process $pp \to t+X \to H^+ b +X$ from $pp \to t+X \to W^+ b  +X$. In particular, single top production 
in hadron colliders is subject itself to backgrounds~\cite{Aad:2009wy} which we have not considsered here.
The most relevant of these backgrounds is the $W b\bar{b}$ production. Needless to say that the $\cos \psi$, the
polarisation information on the $\tau^\pm$ from the decay $\tau^\pm \to \rho^\pm \nu_\tau$, and the transverse
mass distribution will retain their discriminant power to suppress them, albeit at the cost of a small loss in 
the significance of the signal. We plan to take this into account together with a complete treatment of 
the detector effects in a forthcoming more realistic analysis, which is required to assign an error on the
charged Higgs mass due to such effects.

\section{Summary and Outlook}
We have reported here an analysis with improved sensitivity to charged Higgs searches in top quark
decays $t \to bH^+ \to b \tau^+\nu_\tau$ at the LHC. We concentrate on hadronic $\tau^\pm$ decays,
in particular, the decay mode $\tau^\pm \to \rho^\pm \nu_\tau$, and take into account the polarisation
information of the $\tau^\pm$ passed on to $\rho^\pm$. The observables which play a dominant role in
our analysis are the energy and $p_T$ of the $b$-jets from the decays $t \to bW^+$ and $t \to b H^+$,
energy and $p_T$ of the $\tau^\pm$-jets from the two decay chains, and the energy and $p_T$ of the
single-charged prong ($ \pi^\pm$ coming from the decay chain
 $\tau^\pm \to \rho^\pm \nu_\tau \to \pi^\pm \nu_\tau$).  Distributions in these variables are studied
together with angular distribution in $\cos \psi$ defined in eq.~\ref{eq:angle-psi}. This information
is fed to a multivariate analysis using the BDTD techniques. The BDTD response shows that a
clear separation between the $t \to bW^+$ and $t \to bH^+$ decays can be achieved in both the $t\bar{t} X$
pair production and the $t/\bar{t} X$ single top production at the LHC. We have also shown that
 using a transverse mass definition, as suggested in~\cite{Gross:2009wia}, 
the process $pp \to t/\bar{t} X$  allows one to
determine sharp Jacobian peaks for the mass of the $H^\pm$-bosons.
 With the benchmark
integrated luminosity of 10 fb$^{-1}$ at 14 TeV, the light charged Higgs ($m_{H^+} < m_t-m_b )$
can be discovered for all values of $\tan \beta$, where the decay mode $H^\pm \to \tau^\pm \nu_\tau$
is dominant.

 In estimating the quoted significances, we have assumed that the decay $t \to bW^+$
makes up the dominant background. This should be refined by taking into account non-$t$-backgrounds,
such as coming from $(Z,W) + {\rm jets}$.

{\bf Acknowledgements:} We thank Merlin Kole, Theodota Lagouri and Torbjorn Sjostrand for helpful discussions.
This research was partially supported by MICINN (Spain) under contract FPA2008-00601.

\end{document}